\documentclass[prd,aps,letter,twocolumn,floatfix,notitlepage]{revtex4-1}

\usepackage{graphicx,psfrag}
\usepackage{mathrsfs,amsmath,amsfonts,amssymb, bm}
\usepackage{multirow}
\usepackage{comment,hyperref}
\usepackage{float}
\usepackage{algorithm}
\usepackage{algpseudocode} 
\usepackage{setspace} 

\newcommand{\be}{\begin{equation}}
\newcommand{\ee}{\end{equation}}
\newcommand{\bsube}{\begin{subequations}}
\newcommand{\esube}{\end{subequations}}

\def\l{\ell}

\def\bmu{{\bm \mu}}
\def\btheta{{\bm \theta}}

\newcommand{\TEOBResum}{$\text{TEOB}_\text{Resum}$}
\newcommand{\TEOBNNLO}{$\text{TEOB}_\text{NNLO}$}

\usepackage{color}
\definecolor{cyan}{rgb}{0,0.9,0.9}
\definecolor{orange}{rgb}{0.9,0.5,0}
\definecolor{magenta}{rgb}{1,0,1}
\definecolor{purple}{rgb}{0.8,0.4,0.8}
\definecolor{gray}{rgb}{0.8242,0.8242,0.8242}

\begin{document}

\title{Effective-one-body waveforms for binary neutron stars using surrogate models}

\author{
Benjamin D. Lackey$^1$, 
Sebastiano Bernuzzi$^{2, 3}$, 
Chad R.~Galley$^3$, 
Jeroen Meidam$^4$, 
Chris Van Den Broeck$^4$
}
\affiliation{
$^1$Department of Physics, Syracuse University, Syracuse, NY 13244, USA\\
$^2$DiFeST, University of Parma, and INFN, I-43124, Parma, Italy\\
$^3$Theoretical Astrophysics, Walter Burke Institute for Theoretical Physics, California Institute of Technology, Pasadena, California 91125, USA\\
$^4$Nikhef -- National Institute for Subatomic Physics, Science Park 105, 1098 XG Amsterdam, The Netherlands
} 

\date{\today}

\begin{abstract}

Gravitational-wave observations of binary neutron star systems can
provide information about the masses, spins, and structure of neutron
stars. However, this requires accurate and computationally efficient
waveform models that take $\lesssim 1$~s to evaluate for use in
Bayesian parameter estimation codes that perform $10^7 - 10^8$
waveform evaluations. We present a surrogate model of a nonspinning effective-one-body 
waveform model with $\ell = 2$, $3$, and $4$ tidal
multipole moments that reproduces waveforms of binary neutron star numerical simulations up to merger.
The surrogate is built from compact sets of effective-one-body waveform 
amplitude and phase data that each form a reduced basis.
We find that 12 amplitude and 7 phase basis elements are sufficient to reconstruct 
any binary neutron star waveform with a starting frequency of 10~Hz.
The surrogate has maximum errors of 3.8\% in amplitude (0.04\% excluding the last 100M before merger) 
and 0.043 radians in phase. The version implemented in the LIGO Algorithm Library takes $\sim 0.07$~s to evaluate for a 
starting frequency of 30~Hz and $\sim 0.8$~s for a starting frequency of 10~Hz, 
resulting in a speed-up factor of $\sim10^3$--$10^4$ relative to the original Matlab code. 
This allows parameter estimation codes to run in days to weeks rather than years, 
and we demonstrate this with a Nested Sampling run that recovers the masses 
and tidal parameters of a simulated binary neutron star system.

\end{abstract}


\maketitle




\section{Introduction}

One of the primary targets for gravitational-wave detectors such as Advanced LIGO (aLIGO)~\cite{Harry2010}, 
Advanced Virgo~\cite{Acernese2009}, KAGRA~\cite{Somiya2012}, and 
LIGO-India~\cite{IyerSouradeepUnnikrishnan2011} is the inspiral of binary neutron star (BNS) systems. 
The evolution of the waveform provides detailed information about the masses and spins of the two 
neutron stars (NSs) as well as information about the NS structure and equation of state (EOS) 
encoded in the tidal interactions of the two NSs.

Measuring the parameters of the binary, however, requires waveform models that are both fast, 
for use in Bayesian parameter estimation codes, and accurate, to minimize systematic errors 
in the recovered parameters. Almost all previous Bayesian parameter-estimation studies of 
BNS systems~\cite{DelPozzoLiAgathos2013, WadeCreightonOchsner2014, LackeyWade2015, 
AgathosMeidamDelPozzo2015, ChatziioannouCornishKlein2015} have used post-Newtonian (PN) 
waveform models~\cite{DamourIyerSathyaprakash2001, BuonannoIyerOchsner2009}. 
These models typically take $\sim 1$~s or less to evaluate, and are therefore suitable for 
Markov-Chain-Monte-Carlo or Nested-Sampling codes (see~\cite{VeitchRaymondFarr2015} for a review) 
that require $10^6$--$10^7$ waveform evaluations. Unfortunately, because the PN expansion is only 
known completely to 3.5PN order, the uncertainty in the waveform phase can be greater than 
$\sim 10$~radians by the time the binary merges~\cite{HindererLackeyLangRead2010}. 
This can lead to significant biases in the measured masses and tidal 
interactions~\cite{Favata2014, YagiYunes2014, WadeCreightonOchsner2014}.

The effective one body (EOB) formalism first introduced in
Ref.~\cite{BuonannoDamour1999} provides an alternative that includes
several effects beyond the standard PN expansion (see
Ref.~\cite{DamourNagar2010} for a review) and can be calibrated with
numerical relativity binary black hole (BBH) simulations near 
merger~\cite{PanBuonannoBoyle2011, TaracchiniPanBuonanno2012, TaracchiniBuonannoPan2014, 
DamourNagarBernuzzi2013, NagarDomourReisswig2015}. Most implementations of EOB waveforms, 
however, are significantly slower than for PN waveforms, sometimes taking tens of minutes to 
generate a single waveform, and this is unusably slow for most parameter estimation algorithms. 
Recent work on optimizing EOB waveform generation for BBH has resulted in a significant 
speed-up~\cite{DevineEtienneMcWilliams2016}, but this optimization must be done 
for each new waveform model and will not work for numerical relativity simulations.

Reduced-order modeling (ROM) techniques provide a framework for reducing large data sets that can be 
used to build lightweight models that are rapidly evaluated as a substitute, or surrogate, in place 
of the slow waveform generation code. The method, introduced in~\cite{FieldGalleyHesthaven2014}, 
begins with a training set of waveforms that covers the waveform parameter space. 
A greedy algorithm~\cite{Cormen:2001, binev2011convergence, 2012arXiv1204.2290D, Hesthaven2014} exposes the most relevant 
waveforms needed to accurately represent the full training set~\cite{Field:2011mf}. 
These relatively few number of judiciously chosen waveforms, also called a reduced basis, 
captures the dependence of the training set waveforms on parameters.  

Waveforms for arbitrary parameter values (in the training region) can then be generated from the 
reduced basis by estimating the parametric dependence of the projection coefficients.
This can be done in two ways. In one case, each coefficient is interpolated as a function of waveform 
parameters using the training data~\cite{Purrer:2014fza, Purrer:2015tud}. 
In the other case, one uses the empirical interpolation method~\cite{Barrault2004667, chaturantabut2010nonlinear, FieldGalleyHesthaven2014} 
to build an interpolant that is customized to the waveform data such that at a relatively few specific times 
(i.e., the interpolation nodes) one fits for the parametric variation of the waveform data~\cite{FieldGalleyHesthaven2014}. 
This second approach is compact, robust to round-off noise, and allows for the intrinsic waveform errors 
to be incorporated in the error of the final surrogate model (e.g., see~\cite{BlackmanFieldGalley2015}).

Reduced-order modeling in gravitational wave physics started with an observation that the inspiral 
dynamics of precessing BBHs can be dimensionally reduced, meaning that many configurations share 
similar, almost redundant, qualities that vary smoothly across parameter space~\cite{Galley:2010rc}.
The result implied that the multi-dimensional parameter space for precession waveforms might effectively 
be considerably smaller thus providing a possible avenue towards beating the ``curse of dimensionality'' 
for template bank coverage. At the same time, compression factors $\sim 10$ of a small template bank 
of non-spinning gravitational waveforms were achieved using a singular value decomposition~\cite{Cannon:2010qh}.
Subsequently, reduced-order modeling techniques have been used to efficiently represent/compress large 
waveform banks~\cite{Field:2011mf, Cannon:2011xk, Herrmann:2012if, Caudill:2011kv, BlackmanSzilagyiGalley2014} 
and to build fast and accurate surrogate models~\cite{FieldGalleyHesthaven2014} of merger waveforms~\cite{FieldGalleyHesthaven2014, Cannon:2012gq, 
Purrer:2014fza, BlackmanFieldGalley2015, Purrer:2015tud}, which can be used in multiple-query applications 
like parameter estimation studies~\cite{Canizares:2013ywa, Canizares:2014fya, Smith:2016qas} that use reduced-order
quadratures~\cite{antil2013two}. In fact, reduced-order models are crucial in modern GW search pipelines~\cite{Cannon:2011vi} 
and in parameter estimation studies to accelerate waveform generation and likelihood computations.

In this work, we construct a reduced-order surrogate model (or ``surrogate'') 
for the $\ell=m=2$ mode of BNS waveforms generated with the EOB
formalism.  This EOB model, described in Ref.~\cite{Bernuzzi:2014owa},
incorporates tidal interactions that are parameterized by the
quadrupolar $\ell=2$ tidal deformability $\Lambda_2$ of each star as
well as the $\ell=3$ and 4 tidal deformabilities $\Lambda_3$ and
$\Lambda_4$, respectively. These tidal interactions enter at the 5th,
7th, and 9th PN orders, respectively in a resummed form, and lead
to an accumulating phase shift of $\sim 1$~radian up to a
gravitational-wave frequency of 400Hz and $\sim 10$~radians up to
the BNS merger frequency as shown in Fig.~\ref{fig:eosphase} below.
(An alternative model that includes
tidally excited resonances has recently been developed~\cite{HindererTaracchiniFoucart2016}.) We construct 
separate reduced bases for the amplitude and phase, and find them to be extremely compact; 12 amplitude 
bases and 7 phase bases are sufficient to accurately reproduce any waveform in the training set. 
We then interpolate the amplitude and phase as a function of waveform parameters at the times chosen 
by the empirical interpolation method using Chebyshev interpolation. 

Because EOB models with both tidal interactions and spin are just starting to become available, 
our surrogate only applies to nonspinning BNS systems. In future work we intend to incorporate 
NS spins once they are available in the EOB models for tidally interacting systems. 
We note that inspiraling BNS systems are not likely to have significant spins. 
The fastest known NS in a confirmed BNS system has a spin frequency of 44~Hz~\cite{KramerWex2009}, 
corresponding to a dimensionless spin of $\sim 0.04$. Another potential BNS system has a NS with a spin 
frequency of 239~Hz~\cite{LynchFreireRansom2012}, corresponding to a dimensionless spin of $\sim 0.2$.
However, even a spin of $\sim 0.03$ can lead to a systematic bias in the estimated tidal parameters that are
as large as the statistical errors if not incorporated into the waveform model.~\cite{Favata2014, YagiYunes2014}.

We organize the paper as follows. In Section~\ref{sec:teob}, we summarize the EOB model for BNS systems 
from which we construct the reduced-order surrogate model. We describe the steps to build the surrogate in 
Section~\ref{sec:rom}, and present its accuracy and speed for predicting EOB waveforms at new parameter 
values in Section~\ref{sec:results}. Finally, we summarize our results and discuss future work in Section~\ref{sec:theend}. 
In the Appendix, we describe the accuracy of approximating the $\ell=3$ and 4 tidal interactions in terms of the $\ell=2$ tidal interaction.

\textit{Conventions:} Unless explicitly stated, we use units where $G=c=1$.

\section{Tidal EOB waveform model}
\label{sec:teob}

\subsection{\TEOBResum{}}

In this work, we use the tidal EOB (TEOB) model developed in
\cite{Bernuzzi:2014owa} and called \TEOBResum{}. \TEOBResum{} incorporates
an enhanced (resummed) attractive tidal potential derived from recent
analytical advances in the PN and gravitational self-force description of
relativistic tidal interactions \cite{Damour:2009wj,Bini:2014zxa}. 
The resummed tidal potential of \TEOBResum{} significantly improves the
description of tidal interactions near the merger over the previous
next-next-to-leading-order (NNLO) TEOB model \cite{Damour:2009vw,Bernuzzi:2012ci}
and over the conventional PN models. 
In particular, \TEOBResum{} predicts high-resolution, multi-orbit numerical relativity
results within their uncertainties and without fitting parameters to
BNS numerical waveforms \cite{Bernuzzi:2014owa}. 

The main features of \TEOBResum{} are summarised in what follows.
The Hamiltonian is $H_{\rm EOB} = M\sqrt{1+2\nu(\hat{H}_{\rm eff}-1)}$
with 
\begin{align}
 &\hat{H}_{\rm eff}(u,p_{r_*},p_\varphi)\equiv \\ 
&\sqrt{A(u;\nu)\,(1  + p^2_\varphi u^2 + 2\nu(4-3\nu)u^2 p_{r*}^4) +
  p_{r*}^2} \ , \nonumber 
\end{align}
where the binary mass is $M=M_A+M_B$, the
symmetric mass-ratio is $\nu=M_AM_B/M^2$, $u\equiv
1/r$, $r$ is the EOB radial coordinate, and $p_\varphi$ and $p_{r_*}$ are
the conjugate momenta (see e.g. \cite{Damour:2009vw}). The EOB potential 
\be
A(u;\nu)\equiv A_0(u;\nu)+A_T(u;\nu) \ , 
\ee
is the sum of a point-mass term and a tidal term. 
$A_{0}(u;\nu)$ is
defined as the $(1,5)$ Pad\'e approximant of the formal 5PN expression
$A_{0}^{\rm 5PN}(u;\nu)=1-2u +a_3 u^{3} + a_4 u^4 +
(a_{5}^{c}(\nu)+a_{5}^{\ln}\ln u) u^{5} +
(a_{6}^{c}(\nu)+a_{6}^{\ln}\ln u) u^{6}$. 
The coefficients up to 4PN, i.e.~$(a_3,a_4,a_5^c(\nu),a_5^{\ln})$, are
analytically known~\cite{Bini:2013zaa}.  
Although the 5PN term $a_6^{\ln}$ and the linear-in-$\nu$ part
of $a_{6}^{c}(\nu)$ are analytically
known~\cite{Barausse:2011dq,Bini:2013rfa}, 
we use the value $a_6^c(\nu) =3097.3\nu^2 - 1330.6\nu + 81.38$
fit to NR data in \cite{Nagar:2015xqa}.
The tidal term,
\be
\label{AT}
A_T^{(+)}(u;\nu)\equiv - \sum_{\ell=2}^{4} \left[ \kappa^{A}_\l
  u^{2\ell+2}\hat{A}^{(\ell^+)}_A + ({A}\leftrightarrow {B})\right],
\ee 
models the gravito-electric sector of the interaction
\footnote{
  The tidal interactions of a star (e.g.~star A, we omit the labels
  in this footnote) in an external field are
  in general parametrized by
  (i) gravito-electric coefficients
  $G\mu_\l$ $[length]^{2\l+1}$ measuring the $\l$th-order mass
  multipolar moment induced in the star by the external $\l$th-order
  gravito-electric field; 
  (ii) gravito-magnetic coefficients $G\sigma_\l$ $[length]^{2\l+1}$
  measuring the $\l$th-order spin multipolar moment induced in the
  star by the external $\l$th-order gravito-magnetic field; 
  (iii) shape coefficients $h_\l$ measuring the distorsion of the
  surface of the star by an external $\l$th-order gravito-electric 
  field. In the literature, it is customary to consider only the
  dominant gravito-electric interactions (i) and, in most of the cases,
  only the leading order term $\l=2$. The dimensionless Love
  numbers are defined as $k_\l=(2\l-1)!!G\mu_\l/(2R^{2\l+1})$, with
  $R$ the star radius. $\mu_2$ is simply called $\lambda$ in
  \cite{Flanagan:2007ix,Hinderer:2009ca,Vines:2010ca}. Other notations
  are employed in \cite{Vines:2010ca} ($\lambda\to\Lambda$) and in
  \cite{Yagi:2013sva}.}, where 
\bsube
\label{kappal}
\begin{align}
\kappa^A_\ell &= 2 \frac{M_BM_A^{2\l}}{M^{2\l+1}}\frac{k^A_\l}{C_A^{2\l+1}}
= 2 Q^{-1} \left(\frac{X_A}{C_A}\right)^{2\l+1} k^A_\l \ , \\
\kappa^B_\ell &= 2 \frac{M_AM_B^{2\l}}{M^{2\l+1}}\frac{k^B_\l}{C_B^{2\l+1}}
= 2 Q \left(\frac{X_B}{C_B}\right)^{2\l+1} k^B_\l  \ ,
\end{align}
\esube
are the $\l=2,3,4$ tidal polarizability paramaters (or
tidal coupling constants) \cite{Damour:2009wj}.
Labels $A,B$ refer to the stars, $M_{A}$ is the gravitational mass of
star $A$, $R_A$
the areal radius, $C_A=M_A/R_A$, $X_{A}=M_{A}/M$, and $k_\l^A$ are the dimensionless
Love numbers \cite{Hinderer:2007mb,Damour:2009vw,Binnington:2009bb,Hinderer:2009ca}. The expressions above assume $M_A\geq M_B$, so
that $Q=M_A/M_B\geq1$. 
In the equal-mass case, the tidal interaction and EOS information are fully encoded at
leading order (LO) in the total dimensionless 
{\it quadrupolar} tidal coupling constant 
\be 
\kappa^T_2 \equiv \kappa^{(2)}_A + \kappa^{(2)}_B \ .
\ee 
The relativistic correction factors $\hat{A}^{(\ell^+)}_A$ formally include 
all the high PN corrections to the leading-order. The particular choice of 
$\hat{A}^{(\ell^+)}_A$ defines the TEOB model considered in this paper.
The PN-expanded NNLO, fractionally 2PN accurate, expression is 
\be
\label{teob_nnlo}
\hat{A}_A^{(\ell^+ )}(u)= 1+ \alpha^{(\ell)}_1 u+
\alpha^{(\ell)}_2 u^2 \ \ \ {[\rm NNLO]} \ ,
\ee
with $\alpha^{(2),(3)}_{1,2}\neq 0$ computed analytically and
$\alpha^{(4)}_{1,2}=0$~\cite{Bini:2012gu}. This \TEOBNNLO{} model has
been compared against NR simulations in \cite{Bernuzzi:2012ci,Bernuzzi:2014owa},
significant deviations are observed at dimensionless GW frequencies
$M\omega_{22}\gtrsim0.8$, i.e.~after contact and during
the last 2-3 orbits to merger.
The \TEOBResum{} model is defined from \TEOBNNLO{} by substituting the
$\ell=2$ term in \eqref{teob_nnlo} with the expression
\begin{align}
  \label{hatA2}
  \hat{A}^{(2^+)}_A(u) &= 1 + \dfrac{3u^2}{1-r_{\rm LR} u} 
  + \dfrac{X_A \tilde{A}_1^{(2^+) \rm 1SF}}{(1-r_{\rm LR} u)^{7/2}} 
    + \dfrac{X_A^2\tilde{A}_2^{(2^+) \rm 2SF}}{\left(1-r_{\rm
      LR}u\right)^{p}}, 
\end{align}
where the functions $\tilde{A}_1^{(2^+)\rm 1SF}(u)$ and
$\tilde{A}_2^{(2^+)\rm 2SF}(u)$ are given 
in~\cite{Bini:2014zxa} and $p=4$.
The key idea of \TEOBResum{} is to use as pole location in Eq.~\eqref{hatA2}
the light ring $r_{\rm LR}(\nu;\kappa_A^{(\ell)})\,$ of the \TEOBNNLO{}
model, i.e., the location of the maximum of $A^{\rm
  NNLO}(r;\,\nu;\,\kappa_A^{(\ell)})/r^2$. \TEOBResum{} is completed
with a resummed waveform \cite{Damour:2008gu} that includes the NLO tidal
contributions computed
in~\cite{Damour:2009wj,Vines:2010ca,Damour:2012yf}. 

A black hole limit of \TEOBResum{} is given by setting
$\kappa^{(\l)}_{A,B}\to0$. Waveforms obtained this way, however, do
not accurately represent BBH ones because the model does not include
next-to-quasicircular corrections tuned to BBH NR data (so it actually
differs from the model of \cite{Nagar:2015xqa}). For this
reason we exclude the $\kappa^{(\l)}_{A,B}=0$
configurations from the surrogate model. That is not a serious
limitation because BBH waveform models are independently available, and 
because BBH sources are not expected in the mass range covered by our
surrogate. Additionally, including the correct BBH limit would
introduce a discontinuity in the waveform's parameters space that
would affect the overall accuracy of the surrogate.
In a similar way, configurations with $\kappa^{(\l)}_{B}\to0$ approximate
black hole-neutron star binaries, but with an astrophysically
unexpected small mass ratio $Q$. 

The \TEOBResum{} waveform model is determined by seven input parameters (7D
parameter space): the binary
mass-ratio $Q$ and the $\l=2,3,4$ tidal polarizability paramaters (or
tidal coupling constants) $\kappa^{A,B}_\ell$. The latter are linked to the usual multipolar
dimensionless tidal parameters,
e.g. \cite{Damour:2009wj,Yagi:2013sva}. For each star we define 
\be
\label{barlam}
\Lambda^A_\l= \frac{2 k^A_\l }{C^{2\l+1}_A (2\ell-1)!!} \ ,
\ee
which are proportional to the $Q$-independent part of $\kappa^A_\l$,
and correspond to the multipolar quantities called $\bar{\lambda}_{\l}$ in
\cite{Yagi:2013sva}.

\subsection{Approximation of higher order tidal effects}

In constructing a surrogate, it is extremely important to reduce the dimensionality
of the parameter space as much as possible in order to avoid high-dimensional 
interpolation which is often inaccurate and computationally demanding. Fortunately,
Yagi has found tight correlations between the $\ell=2$ tidal parameter
and the $\ell = 3$ and $\ell = 4$ tidal parameters that are nearly independent of the choice of 
EOS for plausible NS EOS models~\cite{Yagi:2013sva}. Yagi then constructed
fits $\Lambda_3^{\rm fit}(\Lambda_2)$ and $\Lambda_4^{\rm fit}(\Lambda_2)$ for the $\ell = 3$ and 4 
tidal parameters in terms of the $\ell = 2$ tidal parameter. This reduces the 7D
parameter space 
\be
(q, \Lambda_2^A, \Lambda_2^B, \Lambda_3^A, \Lambda_3^B, \Lambda_4^A,
\Lambda_4^B) \ ,
\ee
where $q = M_B/M_A \le 1$ to the 3D parameter space 
\be
(q, \Lambda_2^A, \Lambda_2^B) \ .
\ee

We evaluate the systematic uncertainty from these fits 
for 14 different EOS and for NS masses in the range $M\in[0.9M_\odot, M_\text{max}]$, where
$M_\text{max}$ is the maximum mass. 
For these EOS and masses, we find the $\ell=3$ fit $\Lambda_3^{\rm fit}(\Lambda_2)$
results in fractional errors in the range $-0.098 \le \Delta\Lambda_3/\Lambda_3 \le 0.17$
with an average absolute fractional error of $\langle|\Delta\Lambda_3/\Lambda_3|\rangle = 0.04$.
The $\ell=4$ fit $\Lambda_4^{\rm fit}(\Lambda_2)$
results in fractional errors in the range $-0.22 \le \Delta\Lambda_4/\Lambda_4 \le 0.28$
with an average absolute fractional error of $\langle|\Delta\Lambda_4/\Lambda_4|\rangle = 0.08$.
Details are given in the Appendix.

In Fig.~\ref{fig:eosphase}, we show the contribution of each tidal effect to
the phase evolution of the waveform for both the soft EOS SLY~\cite{sly4} and the stiff
EOS MS1b~\cite{ms}. (See \cite{ReadLackey2009} for the naming convention.) The phase shift due to the tidal interactions is usually 
$\sim 10$~radian for the $\ell=2$ interaction, $\sim 1$~radian for the $\ell=3$ interaction, 
and $\sim 0.1$~radian for the $\ell=4$ interaction for frequencies up to the maximum amplitude. 
Also shown is the typical error expected 
in the phase that results from using the fits $\Lambda_3^{\rm fit}(\Lambda_2)$ and $\Lambda_4^{\rm fit}(\Lambda_2)$
instead of the true values of $\Lambda_{3}$ and $\Lambda_{4}$ determined by the EOS.
This error is smaller than the $\ell=4$ tidal effect with an
overall error of $\lesssim 0.01$~radian.

\begin{figure}[htb!]
\begin{center}
\includegraphics[width=\linewidth]{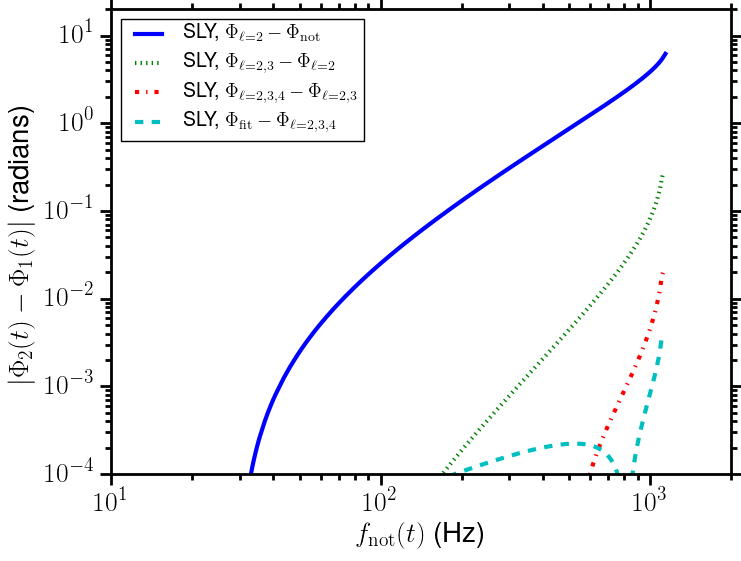}\\
\includegraphics[width=\linewidth]{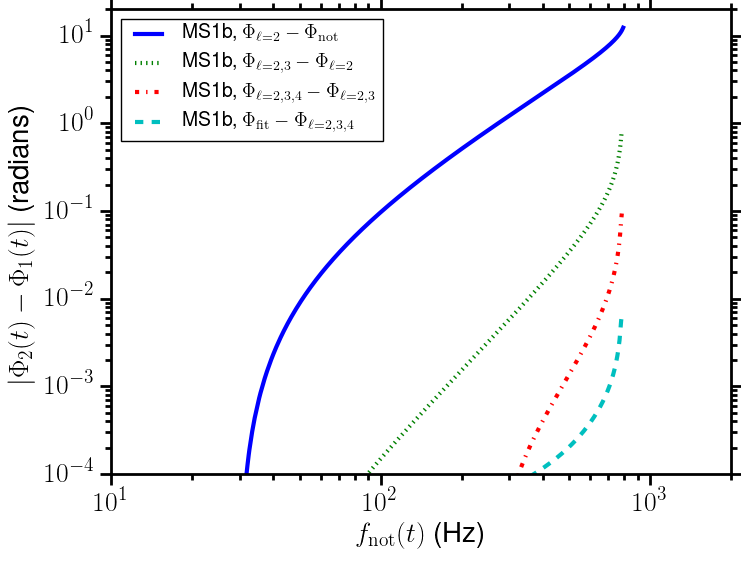}
\caption{Contribution of each tidal multipole moment to the phase evolution of an equal mass 
BNS system with component masses $(M_A, M_B) = (1.4, 1.4)\,M_\odot$ beginning at 30~Hz for the soft EOS SLY (top) and the 
stiff EOS MS1b (bottom). The phase contribution is given by the difference in phase 
between waveforms with no tidal interactions $\Phi_{\rm not}(t)$, only the $\ell=2$ interaction $\Phi_{\ell=2}(t)$, the $\ell=2, 3$
interactions $\Phi_{\ell=2,3}(t)$, and the $\ell=2, 3, 4$ interactions $\Phi_{\ell=2,3,4}(t)$. Also shown by the dashed curve is the error that results from
using the fitting functions $\Lambda_3^{\rm fit}(\Lambda_2)$ and $\Lambda_4^{\rm fit}(\Lambda_2)$ instead of the 
values of $\Lambda_{3}$ and $\Lambda_{4}$ calculated from the EOS $\Phi_{\rm fit}(t)$. 
Each curve is plotted as a parametric function of the phase difference between the two waveforms
$|\Phi_2(t) - \Phi_1(t)|$ versus the frequency of the waveform with no tidal interactions $f_{\rm not}(t)$. 
In this way, the phase difference between waveforms is calculated at
the same time instead of the same frequency. This can be more directly compared to phase errors 
in the surrogate model below which are calculated as a function of time.}
\label{fig:eosphase}
\end{center}
\end{figure}

\section{Building the Tidal EOB waveform surrogate}
\label{sec:rom}

\TEOBResum{} is implemented as a publicly available Matlab code available to download at \cite{eobwebsite}. 
As mentioned above, the time it takes for this code to generate a typical waveform in a ground-based gravitational wave detector's frequency band is about twenty minutes.
Unfortunately, this computational time is far too long for the waveform generation code to be used in practical gravitational wave data analysis applications.
A solution to this problem is provided by reduced-order surrogate modeling, which produces a fast-to-evaluate and compact model that can be used as a substitute for the original \TEOBResum{} code implementation with negligible loss in accuracy.

The recipe for building a reduced-order surrogate model, or surrogate for short, was introduced in~\cite{FieldGalleyHesthaven2014} to which the reader is referred to for further details. 
There are four main steps in the surrogate building process:
\begin{enumerate}
	\setcounter{enumi}{-1}
	\item Precondition the set of precomputed waveforms to vary as smoothly as possible with parameters. This often results in a very compact surrogate model while also improving the accuracy of the surrogate's predictions, in the end;
	\item Build a reduced basis from a set of precomputed waveforms. This results in a compression in parameter space;
		\label{sur:Step1}
	\item Build an empirical interpolant from the reduced basis. This results in a dual compression of the data in the time (or frequency) dimension; 
		\label{sur:Step2}
	\item Estimate or fit for the parametric dependence of the waveform data at specific values of the time (or frequency) samples in the data. 
		\label{sur:Step3}
\end{enumerate}
We discuss the details of these steps for building a surrogate for \TEOBResum{} waveforms in the following subsections. 
However, the set of precomputed waveforms (called a training set) needed for building a reduced basis representation in Step~\ref{sur:Step1}  
often requires preconditioning the data so that the resulting surrogate model will be as compact and accurate as possible. 
This preconditioning step, which can be thought of as the zeroeth step in surrogate building, often involves several choices that must be made in advance, sometimes with input and foresight of steps further down the surrogate building process. 
For example, the size of the reduced basis generated in Step~\ref{sur:Step1} depends crucially on the features and morphologies of the training set waveforms, which can be minimized through the choices made in preconditioning the training data.
As such, building a surrogate may involve a few iterations to converge to the particular strategy and set of choices that end up being suitable for achieving the desired evaluation speed and accuracy. 
In the next subsection, we discuss how the training data of \TEOBResum{} waveforms was generated as well as the choices we made for preconditioning the data.

\subsection{Step 0: Training set and preconditioning}
\label{sec:rom:trains}

When constructing a training set, significant care is required in choosing the waveform parameters as the choice made can impact the accuracy (and sometimes the ability) to accurately estimate or fit the waveform data in Step~\ref{sur:Step3} above. 
We work in units for the mass parameters where the waveform amplitude and time samples are rescaled by the total mass $M = M_A + M_B$ so that a mass-ratio parameter is the only mass parameter needed for the surrogate.
Three common choices are $q = M_B/M_A \le 1$, $Q = M_A/M_B \ge 1$, and the symmetric mass ratio $\nu = M_A M_B / M^2 \le 1/4$. 
We have found that the symmetric mass ratio $\nu$ is a poor choice for \TEOBResum{} waveforms since the amplitude and phase at fixed times can sometimes change rapidly between grid points in the training set of parameters, which will make fitting for the parametric variation (see Step~\ref{sur:Step3} above) difficult. 
The amplitude and phase vary less rapidly as functions of $q$ and $Q$, and we find that $q$ provides slightly better accuracy of the final surrogate model. 
We choose $q\in[0.5, 1]$ as the mass ratio parameter.

For the tidal parameters, we use a rectangular grid of $\Lambda_2^A\in[50, 5000]$ for the more massive NS and $\Lambda_2^B\in[50, 5000]$ for the less massive NS. 
For any realistic EOS, $\Lambda_2^B \ge \Lambda_2^A$, so only a triangular half of this rectangular grid is physically plausible.
However, because most implementations of accurate interpolation algorithms for multi-dimensional data require rectangular grids, we will sample the entire rectangular grid. 
Other choices for the two tidal parameters include $\kappa_2^A$ and $\kappa_2^B$ (defined in Eq.~\eqref{kappal}), as well as $\tilde\Lambda$ and $\delta\tilde\Lambda$, which are another linear combination of $\Lambda_2^A$ and $\Lambda_2^B$ used in parameter estimation~\cite{Favata2014, WadeCreightonOchsner2014}. 
Both of these alternative choices suffer from the same problem.
In particular, mapping a rectangular grid of $\{\kappa_2^A, \kappa_2^B\}$ or $\{\tilde\Lambda, \delta\tilde\Lambda\}$ to the corresponding values of $\{\Lambda_2^A, \Lambda_2^B\}$ can take $\Lambda_2$ outside the domain where the fits $\Lambda_3^{\rm fit}(\Lambda_2)$ and $\Lambda_4^{\rm fit}(\Lambda_2)$ (Eq.~\eqref{eq:lamda34fit}) are valid. 
This makes it impossible to evaluate the $\ell=3, 4$ tidal parameters.

\begin{figure}[t]
\begin{center}
\includegraphics[width=\linewidth]{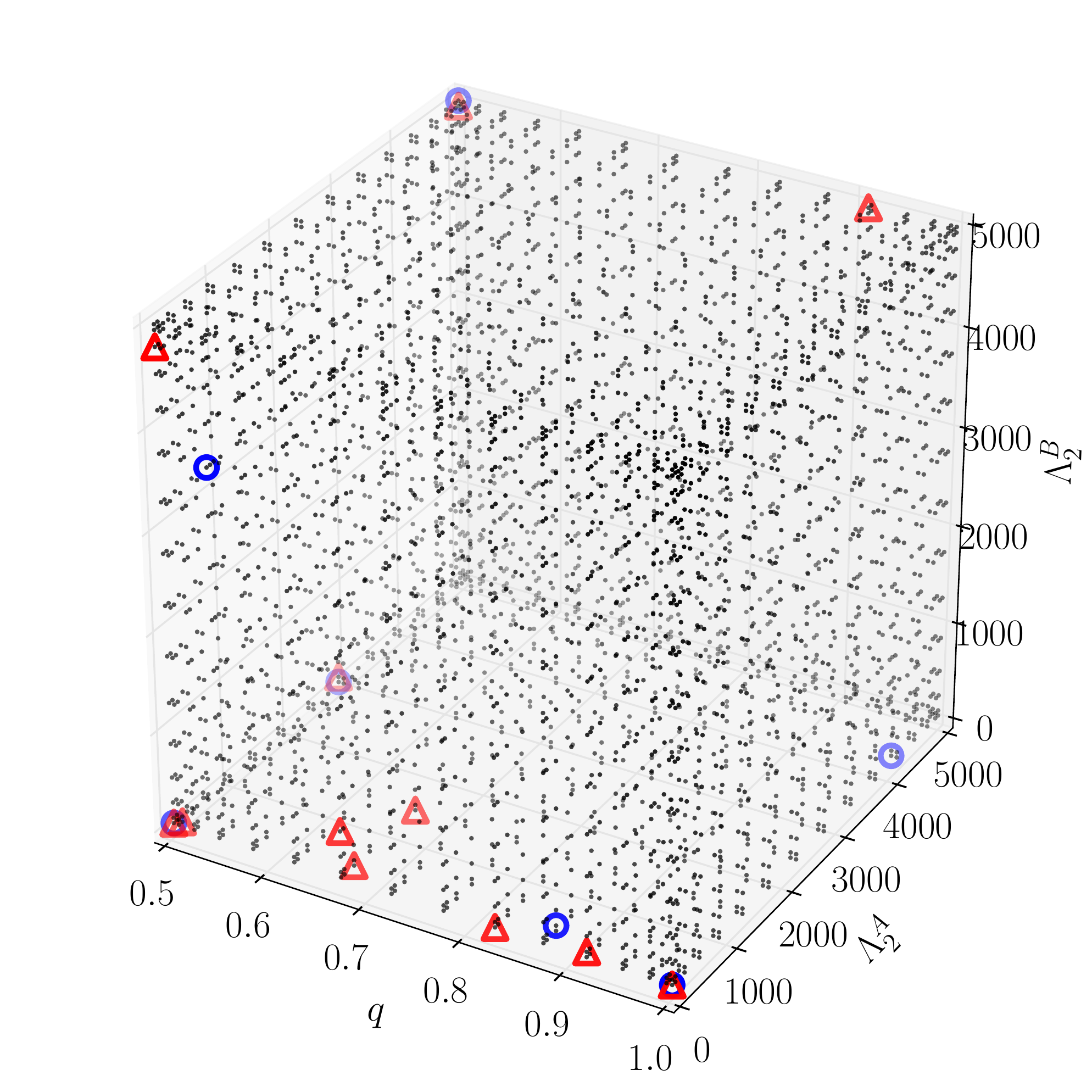}
\caption{The training set is constructed from the Chebyshev-Gauss-Lobatto nodes with 
16 nodes in each dimension for a total of 4096 waveforms. The waveform parameters 
have the range $q\in[0.5, 1]$, $\Lambda_2^A\in[50, 5000]$, and $\Lambda_2^B\in[50, 5000]$. 
The same grid is also used for the Chebyshev interpolation to evaluate the amplitude 
and phase at the empirical nodes $\tau_j$. Red $\triangle$'s represent the 12 waveforms 
chosen by the greedy algorithm to generate the amplitude reduced basis, while blue $\bigcirc$'s 
represent the 7 waveforms chosen for the phase reduced basis.}
\label{fig:trainingset}
\end{center}
\end{figure}

With this choice of parameters, 
\begin{align}
	\btheta := (q, \Lambda_2^A, \Lambda_2^B), 
\label{eq:paramtuple1}
\end{align}
we next choose a discretization of the parameter space to define the training set that will be used in Step~\ref{sur:Step1}. 
As discussed in Sec.~\ref{sec:interpolation} below, we will use Chebyshev interpolation to fit for the variation of the 
amplitude and phase in terms of the waveform parameters $\btheta$. 
So, in constructing our training set, we choose waveform parameters at Chebyshev-Gauss-Lobatto nodes~\cite{Boyd2000}. 
For each parameter, after linearly rescaling the range to $x\in[-1, 1]$, the location of the $M$ nodes are given by
\begin{equation}
\label{eq:nodes}
	x_k = -\cos\left(\frac{k \pi}{M-1}\right),
\end{equation}
where $k=0, \dots, M-1$.
We have found that a grid of $16 \times 16 \times 16 = 4096$ parameters, shown in Fig.~\ref{fig:trainingset}, is 
sufficient to reach the desired accuracy of the final model. These $N=4096$ points define our training set of parameters, 
${\cal T}_N := \{ \btheta_i \} _{i=1}^N$, that is used in building our surrogate model.
This grid is more densely sampled at the edges of the parameter space, which is convenient because 
the algorithm we use for constructing the reduced basis (see Step~\ref{sur:Step1}) tends to choose 
parameters near the boundary of the training set. We then run the \TEOBResum{} code to generate 
4096 waveforms at these training set parameter values with a starting frequency that is less than 
10~Hz for any combination of parameters with $M_A \ge M_B \ge 1M_\odot$, which corresponds 
to a length of $\sim 2 \times 10^8M$ in dimensionless units. This collection of waveforms constitutes our training data.

As discussed in Appendix E of Ref.~\cite{FieldGalleyHesthaven2014}, it is very helpful that the training set waveforms be accurately aligned at maximum amplitude. 
Otherwise, the amplitude and phase will not be smooth functions of the waveform parameters and the resulting surrogate model may not have a compact size, which automatically results in a loss of evaluation speed. 
We do this alignment by (i) densely sampling a waveform near the time of maximum amplitude, (ii) interpolating the amplitude 
with cubic splines, (iii) numerically finding the maximum of the interpolated amplitude, and (iv) shifting the waveform such that $t=0$ corresponds to the maximum amplitude. 
We then set the phase to zero at the common starting time for the shifted waveform.
This alignment procedure is performed for all waveforms in the training data set.

Finally, we resample all of the training data waveforms to reduce the physical memory storage requirements. 
Before $t=-10^3M$, we sample the amplitude and phase uniformly in phase with a spacing of $\Delta\Phi = \pi$. 
After this time we use a uniform in time sampling with a spacing of $\Delta t = 0.1M$ to capture the more 
complicated behavior near merger. This leads to waveforms that only require $\sim 7\times 10^4$ samples 
compared to $\gtrsim 10^8$ samples if we had sampled uniformly in time with sufficient accuracy to capture 
the behavior near merger. Our nonuniform downsampling allows us to store the entire waveform training 
data in $\sim 7$GB instead of many TB. 
More elaborate downsampling strategies exist including one that uses theoretically derived bounds on cubic spline errors to estimate the largest spacing between data points~\cite{Purrer:2014fza} and one that uses a greedy algorithm to select only those data points that are sufficient to recover the full data set up to a requested accuracy by a spline of a given, arbitrary degree~\cite{GalleySchmidt, romSpline}.

\subsection{Step 1: Reduced basis}
\label{sec:rb}

\vspace{-0.5cm}

{\scriptsize
\begin{algorithm}[H]
\caption{Greedy algorithm for reduced basis}
\label{alg:rb}
\begin{algorithmic}[1]
\State {\bf Input:} $ \{ \btheta_i \, , X(\cdot; \btheta_i) \}_{i=1}^N$,  $\epsilon$ 
\vskip 10pt
\State Set $i=0$ and define $\sigma_0 = 1$
\State {\bf Seed choice} (arbitrary):  $\bmu_1 \in {\cal T}$, $e_1 = X(\cdot; \bmu_1)$
\State RB = $\{ e_1 \}$ 
\While{$\sigma_i \ge \epsilon$}
\State $i=i+1$
\State $\sigma_i = \max_{ \btheta \in {\cal T} } \| X(\cdot; \btheta ) - {\cal P}_{i} X(\cdot; \btheta ) \|^2$ 
\State $\bmu_{i+1} = \text{argmax}_{ \btheta \in {\cal T} } \| X(\cdot; \btheta ) - {\cal P}_{i} X(\cdot; \btheta ) \|^2 $  
\State $e_{i+1} = X(\cdot; \bmu_{i+1} ) - {\cal P}_{i} X(\cdot; \bmu_{i+1} )$ (Gram-Schmidt)
\State $e_{i+1} = e_{i+1} / \| e_{i+1} \|$ {\hskip0.725in} (normalization)
\State RB = RB $\cup \, e_{i+1}$
\EndWhile
\State Set $n = i$
\vskip 10pt
\State {\bf Output:} RB = $\{ e_i \}_{i=1}^n$ and greedy points $\{ \bmu_i \}_{i=1}^n$
\end{algorithmic}
\end{algorithm}
}

With the training set of waveforms in hand we now focus on reducing the data to its essential components in both parameters (this subsection) and time (in the next subsection).
The reduction to a compact set of parameters can be achieved by building a reduced basis such that the projection ${\cal P}$ of any training set waveform onto the basis will be indistinguishable from the original waveform up to some tolerance that is specified.
We build a reduced basis using a greedy algorithm~\cite{Cormen:2001, binev2011convergence, 2012arXiv1204.2290D, Hesthaven2014}, which exposes the most relevant parameters in the training set that capture the salient features of the waveform training data.

The greedy algorithm we use is given in~\cite{FieldGalleyHesthaven2014}, to which we refer the reader for further details, and shown in Alg.~\ref{alg:rb}.
The algorithm terminates after $n$ iterations when the projection error is smaller than a specified tolerance $\epsilon$.
The output of the greedy algorithm includes a set of parameter tuples $\{ \bmu_i \}_{i=1}^n$, sometimes called {\it greedy parameters} or {\it greedy points}, and a reduced basis $\{ e_i(t) \} _{i=1}^n$. We use the symbol $X$ as a place holder for a waveform variable.
In Alg.~\ref{alg:rb} and throughout, $|| \cdot || ^2 = \langle \cdot, \cdot \rangle$ is the squared $L_2$ norm where
\begin{align}
	\langle f, g \rangle = \int dt \, f^*(t) g(t)  
\label{eq:overlap1}
\end{align}
is the integral of the product of two generally complex functions $f(t)$ and $g(t)$ and is sometimes called an {\it inner product}.

To make the reduced basis as compact as possible, we follow \cite{FieldGalleyHesthaven2014, Purrer:2014fza, BlackmanFieldGalley2015, Purrer:2015tud} and represent the complex waveforms by their amplitudes and phases instead of real and imaginary parts.
We then build a separate reduced basis for each of the amplitude and phase training data.
This decomposition is made because the amplitude and phase of a waveform have less variation and features in both time and parameters than do the real and imaginary parts of the waveform, which have many oscillations.
The greedy algorithm is sensitive to waveform morphologies and will tend to increase the size of the reduced basis in order to resolve these structures.
As a result, using an amplitude and phase representation for the waveform allows, in our case, for an extremely compact pair of bases. 

We executed the greedy algorithm separately on the amplitudes and the phases of the training set waveforms.
In practice, we use an iterated, modified Gram-Schmidt process~\cite{hoffmann1989iterative}, which is known to be robust against the accumulation of numerical round-off, to generate the orthogonal basis vectors in line 9 of Alg.~\ref{alg:rb}.
We chose relative tolerances of $\epsilon = 10^{-10}$ and $10^{-15}$ for the amplitude and phase, respectively. 
We found that this resulted in extremely compact reduced basis sizes of $n_A = 12$ for amplitudes and $n_\Phi = 7$ for phases.
These 12 and 7 greedy parameters are the minimal amount of information needed to represent the entire training set of \TEOBResum{} waveform amplitudes and phases, respectively, to within our chosen tolerances, as measured with the squared $L_2$ norm. 
Mathematically, if $A$ and $\Phi$ denote the amplitude and phase of a waveform and if $\{e_i^A(t)\}_{i=1}^{12}$ and $\{e_i^\Phi (t) \}_{i=1}^7$ are their corresponding reduced bases then
\begin{subequations}
\label{eq:basis1}
\begin{align}
	A(t; \btheta) \approx {} & \sum_{i=1}^{12} e^A_i(t) c^A_i(\btheta) \\ 
	\Phi(t; \btheta) \approx {} & \sum_{i=1}^{7} e^\Phi_i(t) c^\Phi_i(\btheta) 
\end{align}
\end{subequations}
where the coefficients are
\begin{subequations}
\label{eq:basiscoeffs1}
\begin{align}
	c_i^A(\btheta) & = \langle e_i^A(\cdot) , A(\cdot; \btheta) \rangle \\
	c_i^\Phi(\btheta) & = \langle e_i^\Phi(\cdot) , \Phi(\cdot; \btheta) \rangle  .
\end{align}
\end{subequations}

\begin{figure}[htb!]
\begin{center}
\includegraphics[width=\linewidth]{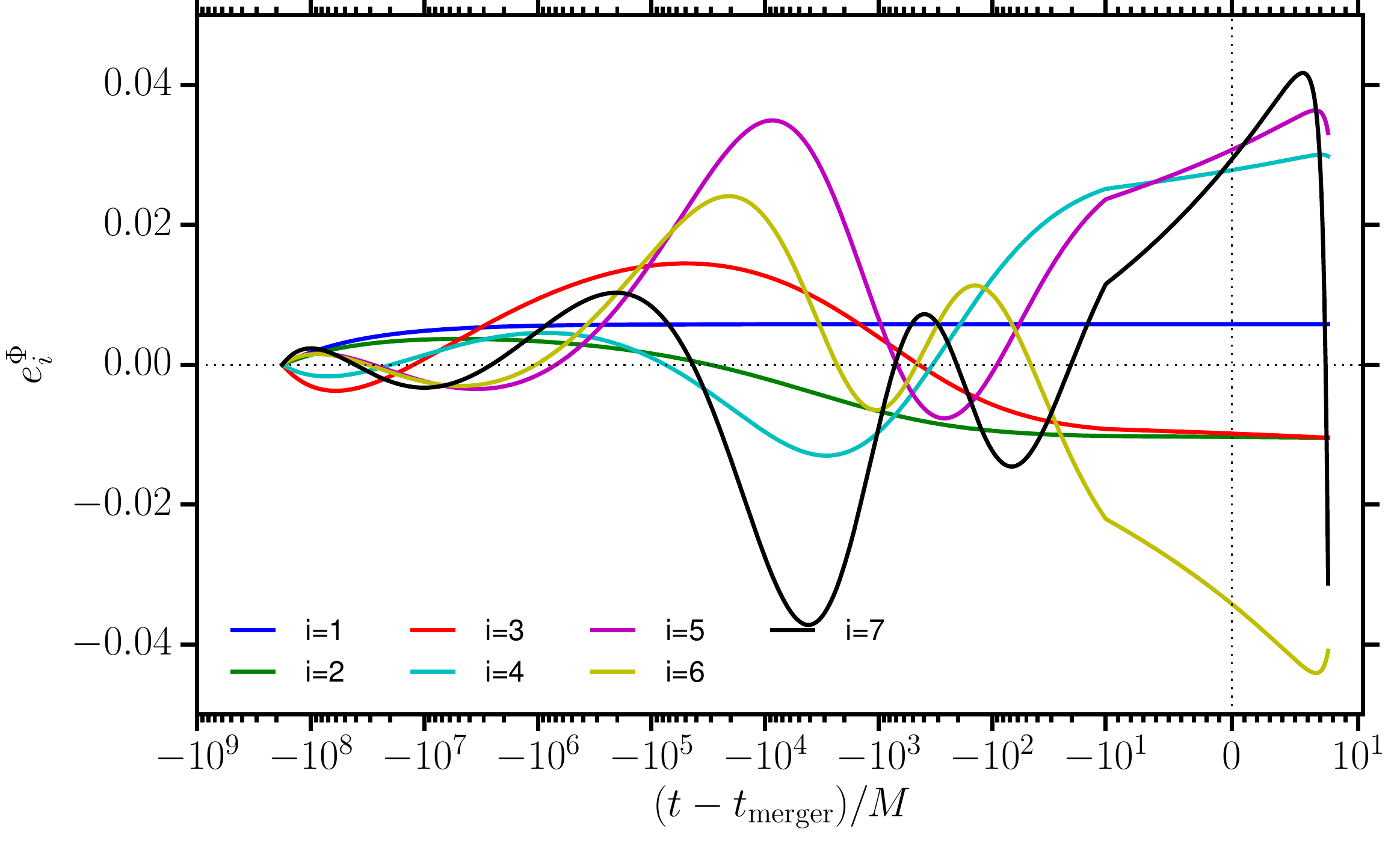}
\caption{The seven orthonormal reduced basis functions $\{e_i^\Phi(t) \}_{i=1}^7$ for the waveform phase. These functions accurately capture the features in the waveform phases within the range of parameters considered. The $x$-axis is linear in the range $[-10^1, 10^1]$ and logarithmic elsewhere.}
\label{fig:RBphase}
\end{center}
\end{figure}

The orthonormal reduced basis elements for the phase are shown in Fig.~\ref{fig:RBphase}.
For comparison, if we had constructed a reduced basis for the complex waveform itself then the basis size would be several hundred to reach a maximum projection error of $10^{-10}$ across the entire training set of waveforms.
The choice to decompose the waveforms into amplitude and phase is thus justified here.

The reduced basis greedy algorithm depends on a choice of seed, which is arbitrarily selected. 
The resulting sequence of parameter tuples selected by the greedy algorithm will depend on that choice of seed.
A different choice still produces a reduced basis that represents all of the training data to within the specified tolerance, by construction.
However, the sizes of the reduced bases built from different seeds tend to vary but will often lie within a few percent of each other so that the seed choice is immaterial~\cite{Field:2011mf, GalleySchmidt}.
What matters is that we have a compact reduced basis to represent accurately the waveform amplitude and phase training data.

Greedy algorithms are extremely flexible for incorporating many decisions and choices for building a reduced-order model.
For example, in a greedy algorithm one can measure the projection errors with the $L_2$ norm, as we did here, or with the $L_\infty$ norm to provide a more stringent requirement of the reduced basis to represent the data in a point-wise sense.
Other problem-specific error measures may be more appropriate (e.g., see~\cite{BlackmanSzilagyiGalley2014}).
One may also implement different greedy algorithms strategies for very large training spaces. 
For example, the training set can be randomly resampled at every iteration of the greedy algorithm, as described in~\cite{BlackmanSzilagyiGalley2014}.
Yet another strategy is to parallelize the computation of the inner product integrals used to compute the projections onto the reduced basis~\cite{greedycpp}.

\subsection{Step 2: Empirical interpolation}
\label{sec:empinterp}

The reduced basis representations of the amplitude and phase in~\eqref{eq:basis1} require knowing the coefficients in~\eqref{eq:basiscoeffs1} or, equivalently, the data $A(t; \btheta)$ and $\Phi(t; \btheta)$ that one is projecting onto the basis.
We wish to predict these coefficients in the linear representations.
One way to do this is to simply fit for the parametric dependence (i.e., $\btheta$) of the coefficients themselves~\cite{Purrer:2014fza, Purrer:2015tud}.
However, the $\btheta$ dependence of the coefficients can become increasingly noisy as the index $i$ increases, which can make the fits unreliable and the ensuing surrogate model evaluations not meet stringent accuracy requirements at new parameter values~\cite{Kaye:thesis, FieldGalleyHesthaven2014}.

The second step of surrogate building aims to provide a sparse subset of times from which it is possible to reconstruct the waveform at any other time by using an {\it empirical interpolant}~\cite{Barrault2004667}, which is informed by the structure and features of \TEOBResum{} waveforms via the reduced basis found in Step~\ref{sur:Step1}.
As we will discuss in the next subsection, fitting for the $\btheta$ dependence of the coefficients is done at each of these time subsamples, which are called {\it empirical interpolation nodes}.
The data being fitted turn out to be robust to the gradual appearance of round-off noise, unlike fitting directly for the projection coefficients mentioned in the previous paragraph~\cite{FieldGalleyHesthaven2014}.

{\scriptsize
\begin{algorithm}[H]
\caption{Empirical Interpolation (EI) Method}
\label{alg:eim}
\begin{algorithmic}[1]
\State {\bf Input:} $\{ e_i \}_{i=1}^n$, $t := \{t_i\}_{i=1}^L$
\vskip 10pt
\State $i = \text{argmax} | e_1(t) |$ (\text{argmax} returns the largest entry of its argument). 
\State Set $T_1 = t_i$
\For{$j = 2 \to n$} 
\State Build ${\cal I}_{j-1} [e_j](t)$ from (\ref{eq:empinterpX1})--(\ref{eq:empinterpX2})
\State $\vec{r} = {\cal I}_{j-1} [e_j](t) -e_j(t)$
\State $i = \text{argmax} |\vec{r}|$
\State $T_j = t_i$
\EndFor
\vskip 10pt
\State {\bf Output:} EI nodes $\{ T_i \}_{i=1}^n$,  interpolant operator ${\cal I}_n$
\end{algorithmic}
\end{algorithm}
}

The algorithm for building an empirical interpolant is given in~\cite{FieldGalleyHesthaven2014}, to which we refer the reader for further details, and shown in Alg.~\ref{alg:eim}.
The empirical interpolant is built from a second greedy algorithm and proceeds as follows~\cite{chaturantabut2010nonlinear}. 
(We focus the presentation on the waveform amplitude for clarity but the same steps are taken for the phase.)
First, we choose a value of time $\tau_1$ from the discrete set of available time samples $t := \{t_i\}_{i=1}^L$ where $L$ is the number of samples.
In our case, we mentioned in Sec.~\ref{sec:rom:trains} that our data have $L = 7 \times 10^4$ time samples that are nonuniformly distributed.
This first time subsample $\tau_1$ is a seed for this greedy algorithm.
In practice, for reasons of conditioning, one chooses the seed to be the time sample at which the first reduced basis function $e_1(t)$ is a maximum in absolute value so that $\tau_1 = {\rm arg~max}_{t} | e_1(t) |$.
We follow this convention here.

For the next step, we build an empirical interpolant.
We label the empirical interpolant ${\cal I}$ of a function $A(t; \btheta)$ by the number $m$ of time subsamples we currently have, namely, ${\cal I}_m[A](t; \btheta)$. 
The notation here is to indicate that ${\cal I}_m$ is an operator that acts on a function $A(t; \btheta)$.
If the function is independent of $\btheta$ then so will the interpolant's operation on that function.

Currently, $m=1$ and we represent the interpolant as a linear combination of the first $m$ reduced basis elements so that
\begin{align}
	{\cal I}_1 [ A ] (t; \btheta) = e_1 (t) C_1 (\btheta) .
\label{eq:eim1}
\end{align}
We assume that the empirical interpolant can always be written in affine form where the dependence on parameters $\btheta$ and time $t$ is factorized.
To solve for the unknown coefficient $C_1(\btheta)$ in~\eqref{eq:eim1} we demand that the interpolant reproduce the data at $t = \tau_1$ so that ${\cal I}_1[A](t; \btheta) = A(t; \btheta)$.
The solution is easily found and given by $C_1(\btheta) =   A(\tau_1; \btheta) / e_1(\tau_1)$ and the empirical interpolant so far is given by
\begin{align}
	{\cal I}_1 [A] (t; \btheta) = B_1(t) A(\tau_1; \btheta)
\label{eq:eim2}
\end{align}
where $B_1(t) = e_1 (t) / e_1 (\tau_1)$. 
In operator form, the $m=1$ empirical interpolant is ${\cal I}_1[\cdot] = B_1 (t) ( \, \cdot \, |_{t=\tau_1})$.

The second empirical interpolant node $\tau_2$ is the time subsample at which the next reduced basis element $e_2(t)$ and its interpolation with the current interpolant ${\cal I}_1 [ e_2](t)$ is largest in absolute value,
\begin{align}
	\tau_2 := {\rm arg~max}_t \big| e_2(t) - {\cal I}_1 [ e_2 ] (t) \big|
\end{align}
Notice that we are choosing the next time subsample in an effort to improve the empirical interpolant's point-wise representation of the reduced basis elements themselves.
Recall that the reduced basis is all that is needed to accurately span the (training set of) waveforms.
The set of nodes is now $\{ \tau_1, \tau_2\}$ and
\begin{align}
	{\cal I}_2 [A] (t; \btheta) = \sum_{i=1}^2 e_i(t) C_i(\btheta) 
\label{eq:eim4}
\end{align}
is the empirical interpolant at this step.
The coefficients are found as in usual interpolation problems.
At the interpolation nodes $\{\tau_1, \tau_2\}$ we require that~\eqref{eq:eim4} equals to the data $A(\tau_i; \btheta)$ and then solve the linear equation
\begin{align}
	\sum_{i=1}^2 V_{ji} C_i (\btheta) = A(\tau_j; \btheta)
\end{align}
where $V_{ji} := e_i (\tau_j)$, which are the elements of a Vandermonde matrix.
After finding the solution, the $m=2$ empirical interpolant is
\begin{align}
	{\cal I}_2 [A] (t; \btheta) = \sum_{j=1}^2 B_j(t) A( \tau_j; \btheta)
\end{align}
where $B_j(t) = \sum_{j=1}^2 e_i(t) (V^{-1})_{ij}$ for $i=1,2$.

This process is repeated until we have used all $n$ of the reduced basis elements to build the final empirical interpolant,
\begin{align}
	{\cal I}_{n} [A] (t; \btheta) = \sum_{j=1}^{n} B_j(t) A(\tau_j; \btheta) 
\label{eq:empinterpX1}
\end{align}
where 
\begin{align}
	B_j(t) = \sum_{i=1}^{n} e_i(t) (V^{-1})_{ij}
\label{eq:empinterpB1}
\end{align}
is the $j^{\rm th}$ element of the empirical interpolation operator ${\bm B}$.
In operator form, the empirical interpolant is
\begin{align}
	{\cal I}_{n} [\cdot]  = \sum_{j=1}^{n} B_j(t) ( \, \cdot \, |_{t=\tau_j} )
\label{eq:empinterpX2}	
\end{align}
Notice that ${\bm B}$ in~\eqref{eq:empinterpB1} is independent of the parameter $\btheta$ and can be computed off-line once the reduced basis is built in Step~\ref{sur:Step1}.
The quantities $\{\tau_j \}_{j=1}^{n}$ are the corresponding interpolation nodes.
In addition, the parameter dependence of the empirical interpolant depends on how the function $A$ varies only at the nodes $\{ \tau_j \}_{j=1}^{n}$.
Finally, notice that, given a parameter tuple $\btheta$, one still needs to know the actual values of the $\{A(\tau_j; \btheta) \}_{j=1}^{n}$ in order to compute the empirical interpolant in~\eqref{eq:empinterpX1}.
We will show how surrogate modeling addresses this issue in Sec.~\ref{sec:interpolation}.

\begin{figure}[htb!]
\begin{center}
\includegraphics[width=\linewidth]{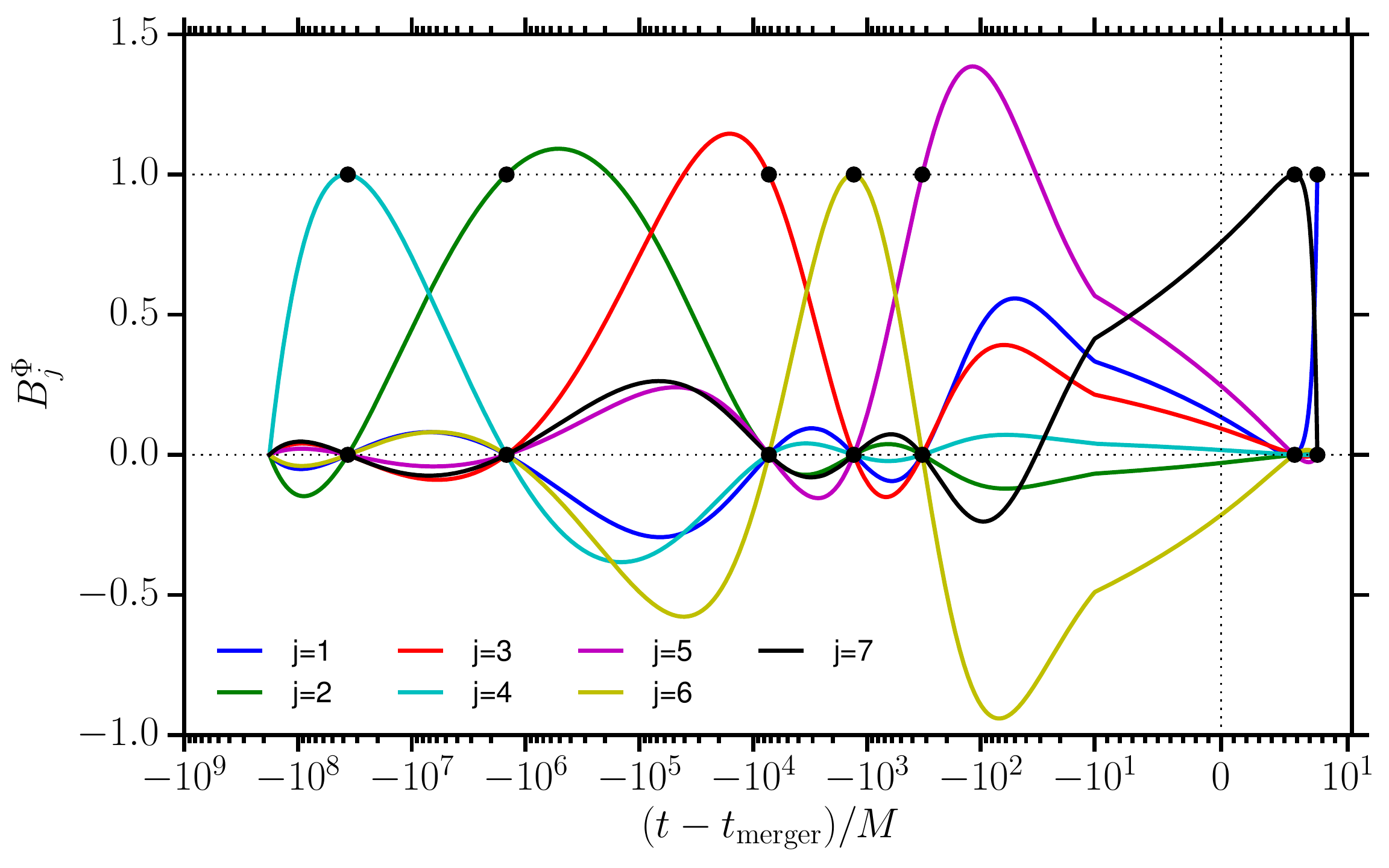}
\caption{The seven elements of the empirical interpolation operator $\{B_j^{\Phi}(t)\}_{j=1}^7$ for the waveform phase. 
The empirical interpolation nodes $\{\tau^\Phi_j\}_{j=1}^{7}$ are shown as black dots.
The $x$-axis is linear in the range $[-10^1,10^1]$ and logarithmic elsewhere.}
\label{fig:Bphase}
\end{center}
\end{figure}

We applied this greedy algorithm to build an empirical interpolant separately for the waveform amplitude and phase, which are given by
\begin{subequations}
\label{eq:empinterpFinal}
\begin{align}
	{\cal I}_{12}^A [ A ] (t; \btheta ) & = \sum_{j=1}^{12} B_j^A(t) A(\tau^A_j; \btheta) \\
	{\cal I}_{7}^\Phi [ \Phi ] (t; \btheta ) & = \sum_{j=1}^{7} B_j^\Phi(t) \Phi(\tau^\Phi_j; \btheta)
\end{align}
\end{subequations}
The interpolating functions for the phase are shown in Fig.~\ref{fig:Bphase}. 
The empirical interpolation nodes $\{\tau^\Phi_j \}_{j=1}^{7}$ are not uniformly spaced because they depend on the underlying features of the \TEOBResum{} waveform family.

The empirical interpolation greedy algorithm we just described and used in this paper is not optimized for speed.
As discussed in Appendix B of~\cite{antil2013two}, the original (discrete) empirical interpolation method algorithm proposed in~\cite{chaturantabut2010nonlinear}, which is the one we discussed above, has a computational cost at the $m^{\rm th}$ step of the greedy algorithm that scales as ${\cal O}(m^4)$.
However, this relatively slow evaluation time is immaterial for our surrogate since the sizes of the amplitude ($n_A =12)$ and phase ($n_\Phi=7$) reduced bases are very small.
A faster algorithm was put forward in \cite{antil2013two} (see Algorithm 5 in their Appendix A) and was used in describing the reduced-order surrogate modeling strategy in~\cite{FieldGalleyHesthaven2014}.
This faster implementation has a cost that scales as ${\cal O}(m^3)$~\cite{antil2013two}.
The computational savings with this faster algorithm is particularly useful for problems involving a large reduced basis, such as the reduced-order model built in~\cite{Smith:2016qas} for the IMRPhenomPv2 waveform family~\cite{Hannam:2013oca} that contained at most $1253$ basis elements.

\subsection{Step 3: Estimating the parametric variation}
\label{sec:interpolation}

The last step in building the surrogate model is to estimate the $\btheta$ dependence at each of the 
empirical interpolation nodes for both the waveform amplitude and phase data. Because the parameter 
space is three dimensional, we intentionally selected our training points in Sec.~\ref{sec:rom:trains} to 
correspond to the nodes of Chebyshev interpolation in three dimensions. Chebyshev interpolation, 
which for smooth $C^\infty$ functions, has errors that converge exponentially with the number of 
Chebyshev polynomials $T_n(x)$~\cite{Boyd2000}. The amplitude and phase at each empirical 
interpolation node are approximated as a tensor product of Chebyshev polynomials,
\begin{subequations}
\label{eq:chebinterp}
\begin{align}
	A(\tau_j^A; \btheta) \approx \tilde{A}_j(\btheta) &:= \sum_{l,m,n} a_{j, lmn}T_l(q)T_m(\Lambda_2^A)T_n(\Lambda_2^B),\\
	\Phi(\tau_j^\Phi; \btheta) \approx \tilde{\Phi}_j (\btheta) &:= \sum_{l,m,n} b_{j, lmn}T_l(q)T_m(\Lambda_2^A)T_n(\Lambda_2^B) .
\end{align}
\end{subequations}
for $j = 1, \ldots n_X$ and $X = \{A, \Phi\}$. Although it would be possible to optimize the number of coefficients 
for each node $\tau_j$, for simplicity we use all $16\times16\times16$ coefficients at each node. The summations 
are efficiently performed using Clenshaw summation~\cite{PressTeukolskyVetterling3rd2007}. The coefficients 
$a_{j, lmn}$ and $b_{j, lmn}$ of the Chebyshev series are precomputed from the known amplitudes and phases 
on the training set grid using Gaussian quadrature~\cite{Boyd2000}. This quadrature is efficiently performed using 
a type-I discrete cosine transform~\cite{scipy:dct}. 

The required fractional accuracy of the interpolation is the maximum allowed amplitude or phase error divided by 
the range of values that the amplitude or phase takes at each empirical node $\tau_j$ over the training set. This 
is most stringent for the phase where, near the merger, we would like the error to be $\lesssim 0.1$~radians but 
the spread is $\sim10^3$~radians over the considered parameter space. This requires a fractional accuracy of 
$\sim10^{-4}$ for the interpolation. 
We assess the errors in the surrogate model due to interpolation in Sec.~\ref{sec:accuracy}.

A distinct advantage of reduced-order surrogate models is that the data output from a simulation or code is used 
{\it directly} for building the model. The first steps (building a reduced basis and and empirical interpolant) 
are accomplished off-line using {\it only} the training data generated by the \TEOBResum{} code. The only model 
inputs come in the last step when we estimate the parameter dependence of the data at the empirical interpolation 
nodes because we implement a choice of fitting functions that can affect the resulting quality of the surrogate predictions.
This often results in a surrogate that nearly retains the accuracy of the underlying training data used to build the model.
In addition, propagating the training data uncertainties and assessing the surrogate errors is fairly straightforward 
because of the minimal amount of modeling inputs, which are isolated to Step~\ref{sur:Step3}. An example of this is 
given in~\cite{BlackmanFieldGalley2015} where a reduced-order surrogate model is built for the gravitational waveforms 
of non-spinning BBH coalescences produced by numerical relativity simulations.

\subsection{Surrogate waveform evaluation}
\label{sec:alldone}

After Step 3, the surrogate model for the amplitude and phase is defined by evaluating the corresponding empirical 
interpolant in~\eqref{eq:empinterpFinal} using the parametric estimations in~\eqref{eq:chebinterp} to predict the values 
at any new parameter values,
\begin{subequations}
\label{eq:surrogate1}
\begin{align}
	A_S (t; \btheta) := \sum_{i=1}^{12} B_i^A(t) \tilde{A}_i (\btheta)  \\
	\Phi_S(t; \btheta) := \sum_{i=1}^7 B_i^\Phi (t) \tilde{\Phi}_i (\btheta)
\end{align}
\end{subequations}
The surrogate evaluation is performed online once a parameter tuple $(q, \Lambda_2^A, \Lambda_2^B)$ is given.

In practical gravitational wave data analysis applications, one
rescales from geometric units to physical units of time and amplitude. 
This rescaling also depends on the total mass $M$ of the compact binary.
In addition, one specifies a starting frequency $f_{\rm start}$. 
The starting time $t_{\rm start}$ corresponding to $f_{\rm start}$ is calculated by numerically solving $f_{\rm start} = f(t_{\rm start})$, where $f(t)=(d\Phi_S/dt)/(2\pi)$ is the frequency once given a tuple of parameter values $(q, \Lambda_2^A, \Lambda_2^B)$. 
We then perform the necessary time and phase shifts and resample the amplitude and phase surrogate predictions in~\eqref{eq:surrogate1}. 
The final expression for the surrogate model waveforms takes the form
\begin{subequations}
\label{eq:surrogate}
\begin{align}
	h_{+S}(t; \btheta) = {} & \frac{1}{2}(1+\cos\iota) \frac{GM}{c^2d} A_S \!\! \left(\frac{c^3t}{GM}; \btheta \right) \nonumber \\
		& \times \cos \Phi_S \!\! \left(\frac{c^3t}{GM} ; \btheta \right), \\
	h_{\times S}(t; \btheta) = {} & \cos\iota \frac{GM}{c^2d} A_S \!\! \left(\frac{c^3t}{GM} ; \btheta \right) \sin \Phi_S \!\! \left(\frac{c^3t}{GM} ; \btheta \right),
\end{align}
\end{subequations}
where $\iota$ is the binary inclination angle and $d$ is the distance to the gravitational wave source.

In this paper, we have built a surrogate model for the amplitude and phase of \TEOBResum{} waveforms in the time domain. 
This surrogate approximates the waveform in a nonlinear representation (because the dependence on phase is nonlinear) and, as such, cannot be used directly for speeding up likelihood computations in parameter estimation studies~\cite{Canizares:2013ywa, Canizares:2014fya, Smith:2016qas} that use reduced-order quadratures~\cite{antil2013two}.
However, it is straightforward to use our surrogate to build a new surrogate for this purpose in the following way.
As we will discuss in the next section, our surrogate accurately predicts waveforms output by the TEOBresum code.
As such, we may use the surrogate above to generate a new training set in $h_+$ and $h_\times$ form, which is a linear representation. 
Then, one can repeat the surrogate building Steps~\ref{sur:Step1}--\ref{sur:Step3} for the two waveform polarizations directly. 
As mentioned earlier, this would generate reduced basis sizes that contain a few hundred elements and so would be somewhat larger in physical memory size and slower to evaluate the resulting surrogate.
Furthermore, one could do this in the frequency domain by computing Fourier transforms of the time-domain training waveforms generated by the surrogate in~\eqref{eq:surrogate} so that the transformation is part of the offline stage.
One may still use our time domain surrogate presented here for parameter estimation studies but we expect significant speed-ups could be obtained by following the strategy just outlined for likelihood computations.

\section{Results}
\label{sec:results}

\subsection{Accuracy}
\label{sec:accuracy}

The required accuracy of the surrogate is determined by the smallest effect that we want to model. 
For BNS systems, this is the tidal interaction that effects the waveform by $\sim 10$~radians up to 
merger for typical EOSs. This means that we will require the waveform error to be significantly 
smaller than $\sim 10$~radians.

\begin{figure}[htb!]
\begin{center}
\includegraphics[width=\linewidth]{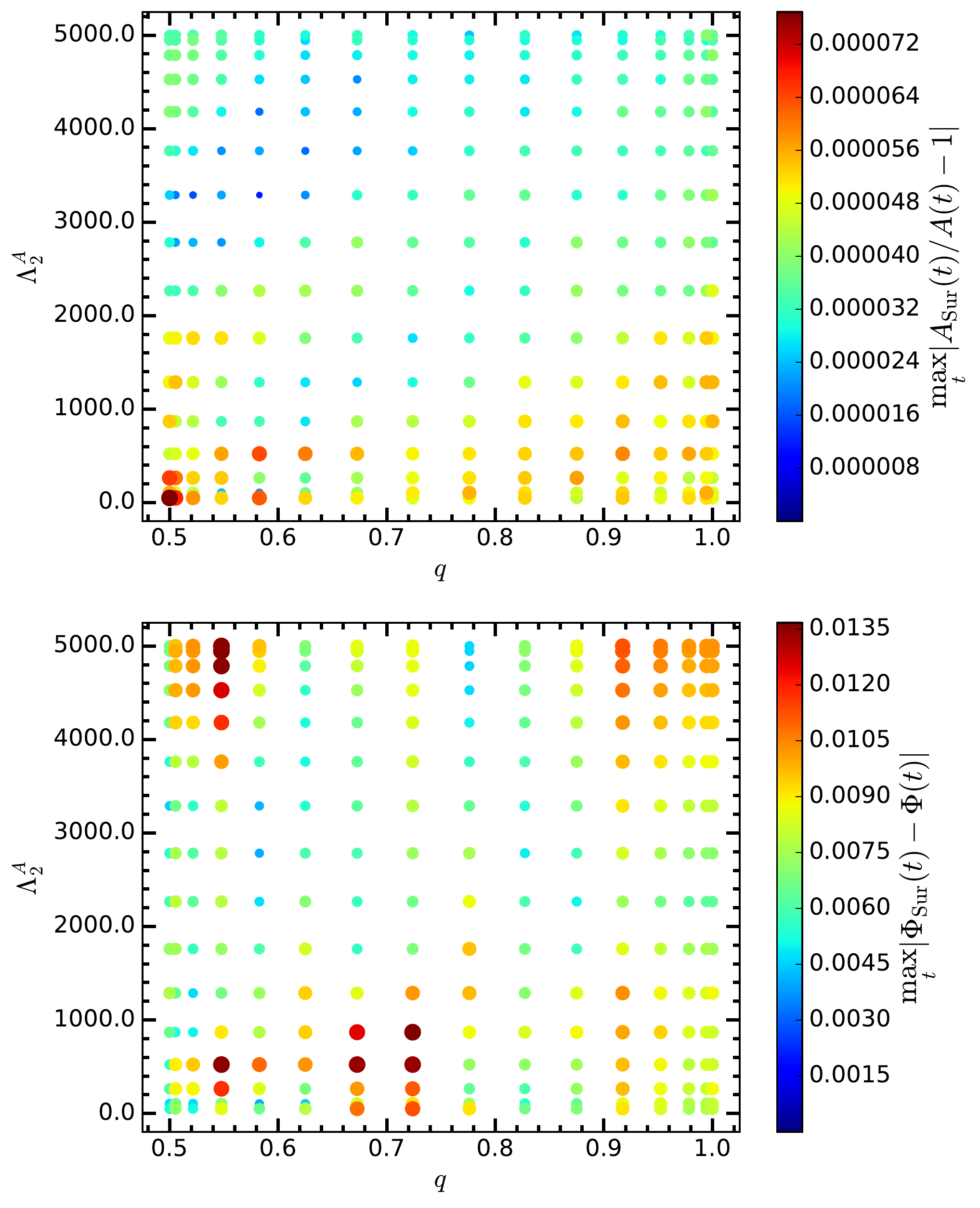}
\caption{Error between the surrogate and the $16^3$ training set waveforms used to construct the 
reduced basis. Larger points represent larger errors. The fractional amplitude and phase 
errors, maximized over time and waveform parameters, are $\Delta A/A = 7.7\times 10^{-5}$ and 
$\Delta\Phi=0.014$ respectively. These errors are due to the finite number of reduced bases.}
\label{fig:errorRB}
\end{center}
\end{figure}

The accuracy of the surrogate can be assessed by comparing it to the training set used to construct 
the reduced basis as well as to a large set of waveforms with randomly sampled parameters. 
Fig.~\ref{fig:errorRB} shows the fractional error in the amplitude as well as the error in the phase 
between the surrogate waveform and each of the 4096 training set
waveforms. The difference between the surrogate and each training set waveform is maximized over all times.
For clarity we suppress the third paramater $\Lambda_2^B$ and only show the maximum error for the 16 values of 
$\Lambda_2^B$ at each grid point. The maximum fractional 
error in amplitude is $\Delta A/A = 7.7\times 10^{-5}$, and the maximum error in phase is $\Delta\Phi=0.014$~radians. 
Because the interpolated values of $A$ and $\Phi$ exactly match the training set at each grid point 
at the empirical nodes $\tau_j$, we see from~~\eqref{eq:surrogate1} that the error in reproducing 
each training set waveform is due almost entirely to the finite number of reduced bases. 
 
\begin{figure}[htb!]
\begin{center}
\includegraphics[width=\linewidth]{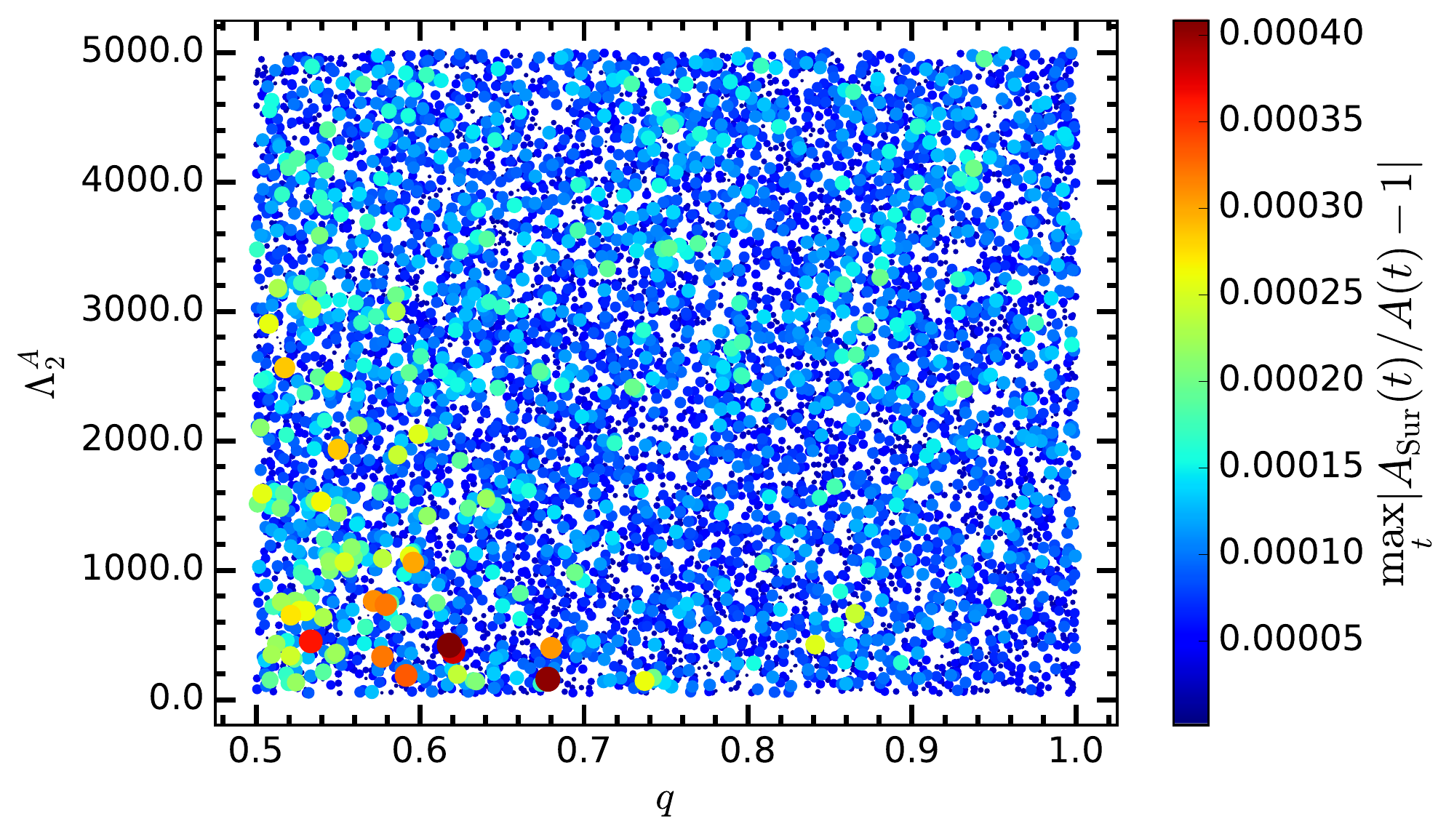}
\includegraphics[width=\linewidth]{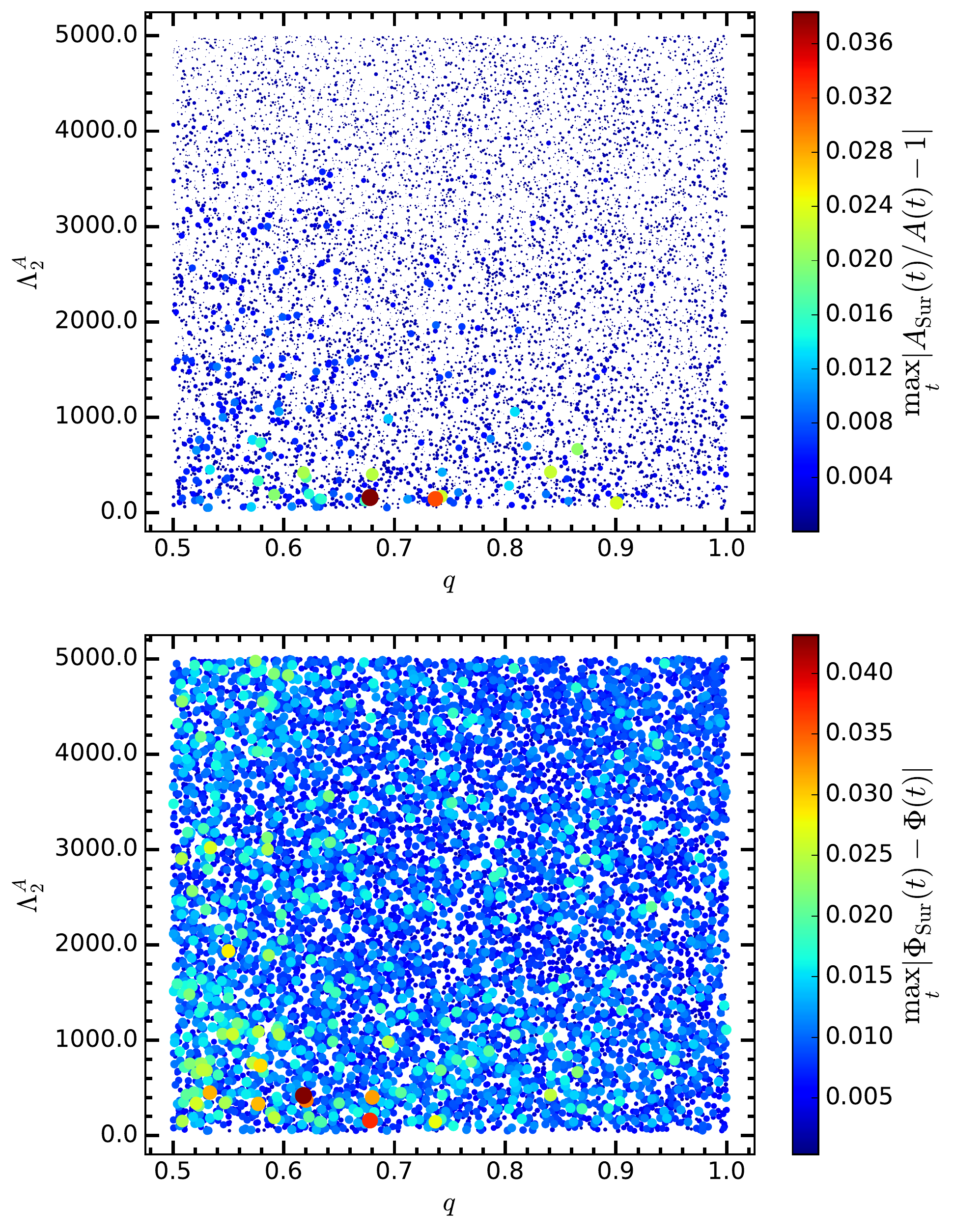}
\caption{Error between the surrogate and $10^4$ waveforms with randomly sampled parameters not 
in the training set. Larger points represent larger errors. Top: Fractional amplitude error $\Delta A/A$ 
maximized over all times except the last $100M$. Middle: Fractional amplitude error $\Delta A/A$ 
maximized over all times. Bottom: Phase error $\Delta\Phi$ maximized over all times.}
\label{fig:errorROM}
\end{center}
\end{figure}

To determine how well the surrogate reproduces a generic waveform within the parameter space, 
we produce $10^4$ waveforms with parameters randomly sampled in the range $q\in[0.5, 1]$, 
$\Lambda_2^A\in[50, 5000]$, and $\Lambda_2^B\in[50, 5000]$. The fractional amplitude and phase 
errors in reproducing the generic waveforms are shown in Fig.~\ref{fig:errorROM}. The errors are 
maximized over time for each waveform. We find a maximum fractional amplitude error of 
$\Delta A/A = 0.038$, but note that, prior to the last $100M$ before merger, the maximum error is
$\Delta A/A = 4\times 10^{-4}$, about two orders of magnitude smaller. The maximum phase error 
is $\Delta\Phi=0.043$~radians.  For both the amplitude and phase, the error is largest for small values 
of the parameters. This results because both the amplitude and phase vary most rapidly for small 
values of the waveform parameters, so the interpolation of the amplitude $A(\tau^A_j; q, \Lambda_2^A, \Lambda_2^B)$ and phase 
$\Phi(\tau^\Phi_j; q, \Lambda_2^A, \Lambda_2^B)$ at each empirical node $\tau_j$ is least accurate there. 

\begin{figure}[htb!]
\begin{center}
\includegraphics[width=\linewidth]{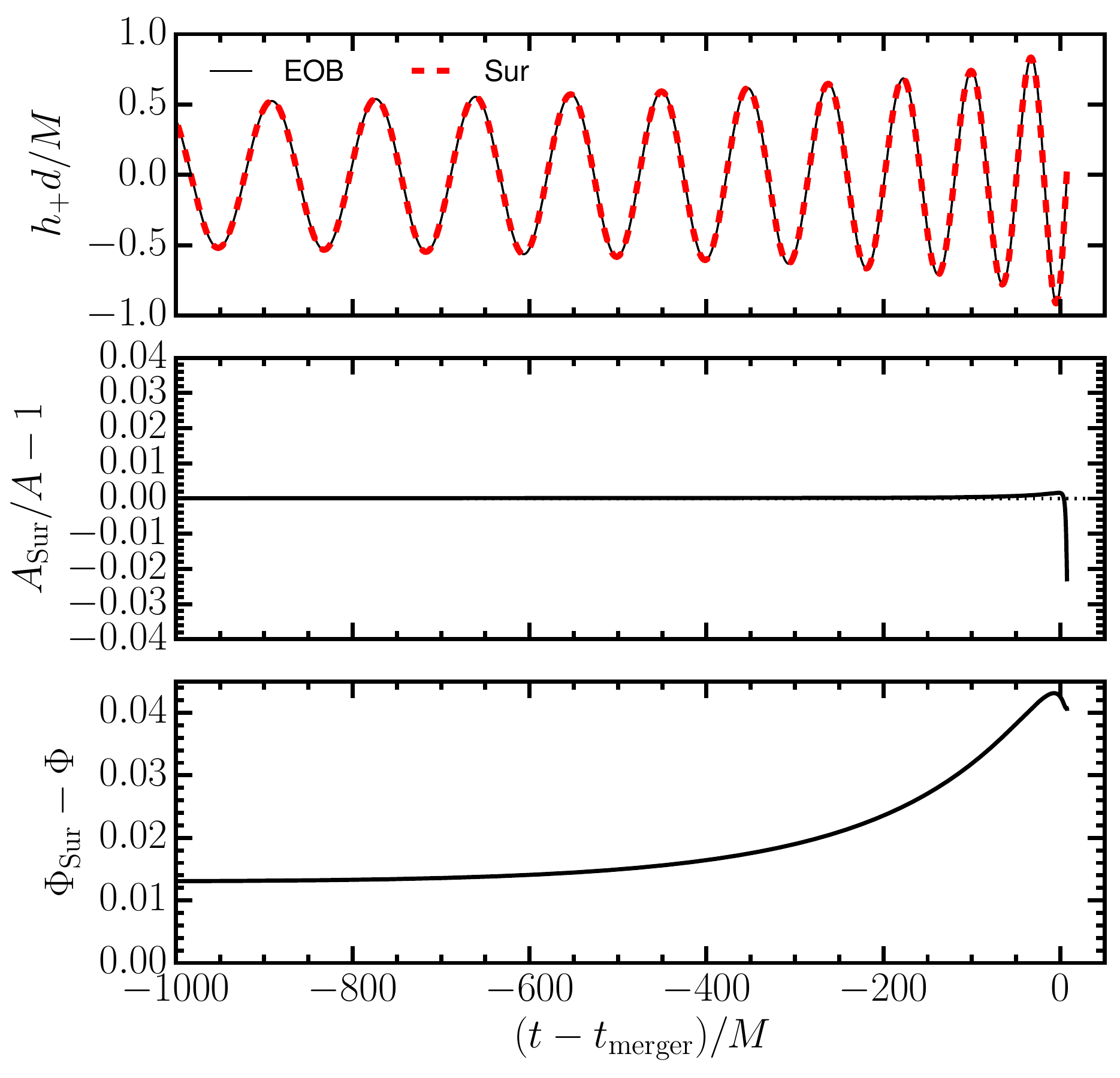}
\caption{Final $\sim 10$ cycles of the waveform with the largest phase error. The parameters are 
$\{q, \Lambda_2^A, \Lambda_2^B\} = \{0.618, 420, 81\}$. The amplitude errors are always largest 
during the last gravitational wave cycle, and are about two orders of magnitude smaller before the 
last gravitational wave cycle.}
\label{fig:errorwave}
\end{center}
\end{figure}

In Fig.~\ref{fig:errorwave}, we show the last $10^3M$ of the $\sim 10^8M$ long waveform that is 
reproduced by the surrogate with the largest phase error. The phase error typically increases gradually 
with time. On the other hand, the amplitude error increases dramatically during the last cycle, and is 
typically two orders of magnitude smaller before the last cycle. This increase in error near the maximum 
amplitude likely results from the finite accuracy with which the training set waveforms are numerically 
aligned at maximum amplitude where the amplitude changes more rapidly with time, making accurate 
interpolation more difficult. 

As a final check of the surrogate model accuracy, we examine the mismatch between 
the surrogate model and the original EOB waveform.
The mismatch represents the loss in signal-to-noise ratio that would result 
from using the surrogate model instead of the original EOB waveform. 
It is defined by the deviation from a perfect overlap after aligning the two waveforms
using the time and phase free parameters $t_0$ and $\phi_0$:
\begin{equation}
\mathcal{M} = 1 - \max_{t_0, \phi_0} \frac{(h_{\rm EOB}, h_{\rm Sur})} {\sqrt{(h_{\rm EOB}, h_{\rm EOB}) (h_{\rm Sur}, h_{\rm Sur})}}.
\end{equation}
The inner product here  is the integral of the Fourier transformed waveforms $\tilde h(f)$ weighted by the noise power spectral 
density (PSD) $S_n(f)$ of the detector:
\begin{equation}
(h_1, h_2) = 4 \Re \int_{f_{\rm low}}^{f_{\rm high}} \frac{\tilde h_1(f) \tilde h^*_2(f)} {S_n(f)} df.
\end{equation}

\begin{figure}[htb!]
\begin{center}
\includegraphics[width=\linewidth]{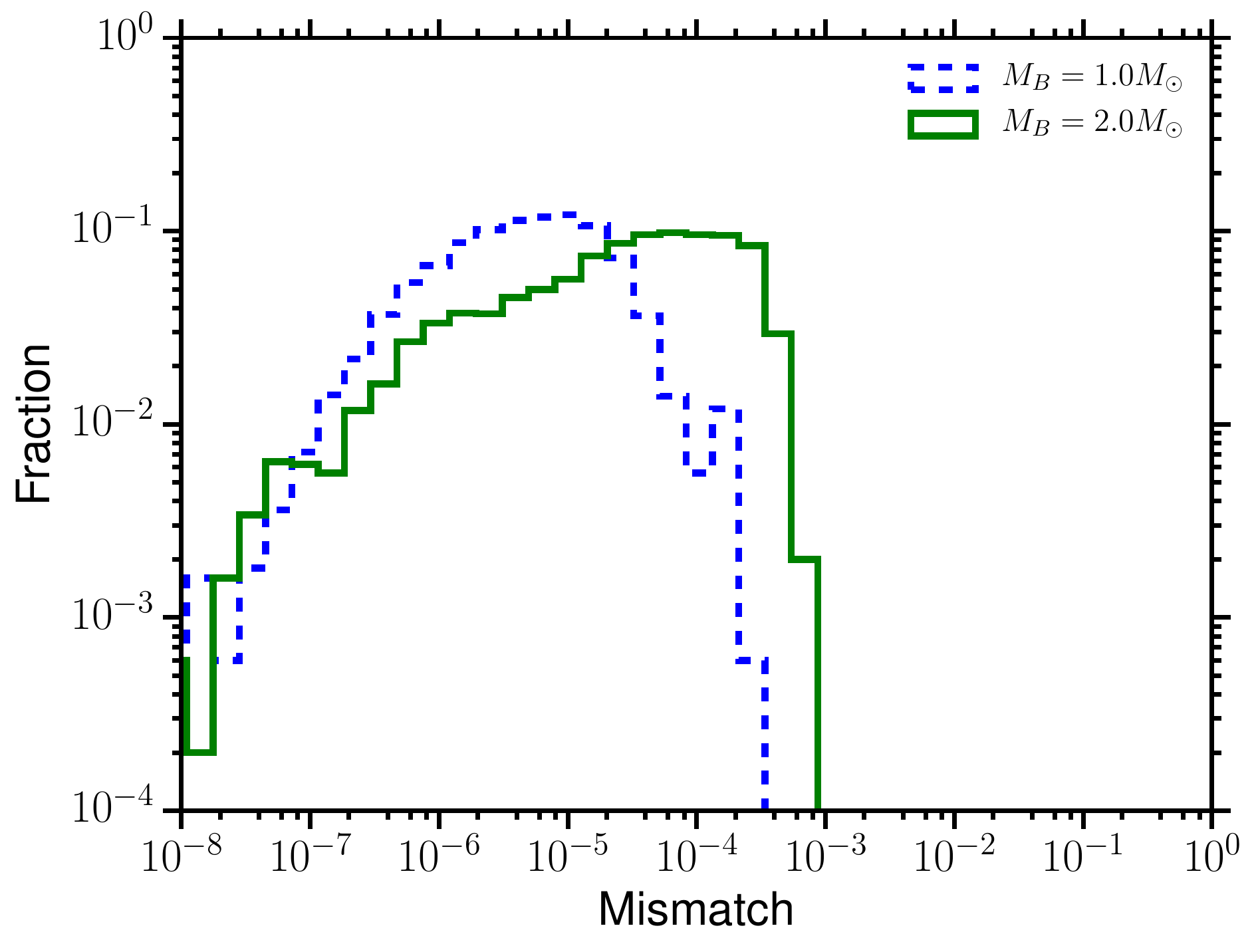}
\caption{Histogram of the mismatch between the surrogate and the $10^4$ randomly sampled EOB waveforms. 
The smaller mass $M_B$ is set to either $1M_\odot$ or $2M_\odot$. The PSD corresponds to the aLIGO
design senistivity and the lower and upper frequency integration bounds are 
$f_{\rm low} = 30$~Hz and $f_{\rm high} = 2048$~Hz.}
\label{fig:mismatch}
\end{center}
\end{figure}

In Fig.~\ref{fig:mismatch}, we show the distribution of mismatch $\mathcal{M}$ between our surrogate
and the $10^4$ randomly sampled EOB waveforms. We use the design sensitivity aLIGO PSD~\cite{Aasi:2013wya} and
a sampling rate of 4096~Hz. Our integration bounds are $f_{\rm low} = 30$~Hz and the Nyquist frequency
$f_{\rm high} = 2048$~Hz. Because the surrogate can be rescaled with mass,
we show results for the smaller mass $M_B$ fixed at $1M_\odot$ or fixed at $2M_\odot$.
The mismatch is larger for the higher mass systems because the frequency where the waveform ends
scales inversely with the total mass, resulting in the less accurate end of the waveform occurring
at smaller frequencies where the detector is more sensitive. Overall, the mismatch is typically smaller than
$\sim 10^{-4}$ except for systems with large component masses, and the mismatch never exceeds $7\times 10^{-4}$.

\subsection{Timing}

One of the main purposes for generating a surrogate is to make parameter estimation more computationally 
efficient. In particular, parameter estimation codes, usually based on Markov Chain Monte Carlo or 
Nested Sampling, typically require $10^7$--$10^8$ sequential waveform evaluations, and the computational time 
is dominated by the waveform evaluation time even for PN waveforms~\cite{Berry2015, Farr2016}. Recently, a few 
parallel algorithms have become available~\cite{ForemanMackeyHogg2013, PankowBradyOchsner2015}. 
However, the performance of these algorithms still scales linearly with the performance of the waveform generator.

\begin{figure}[htb!]
\begin{center}
\includegraphics[width=\linewidth]{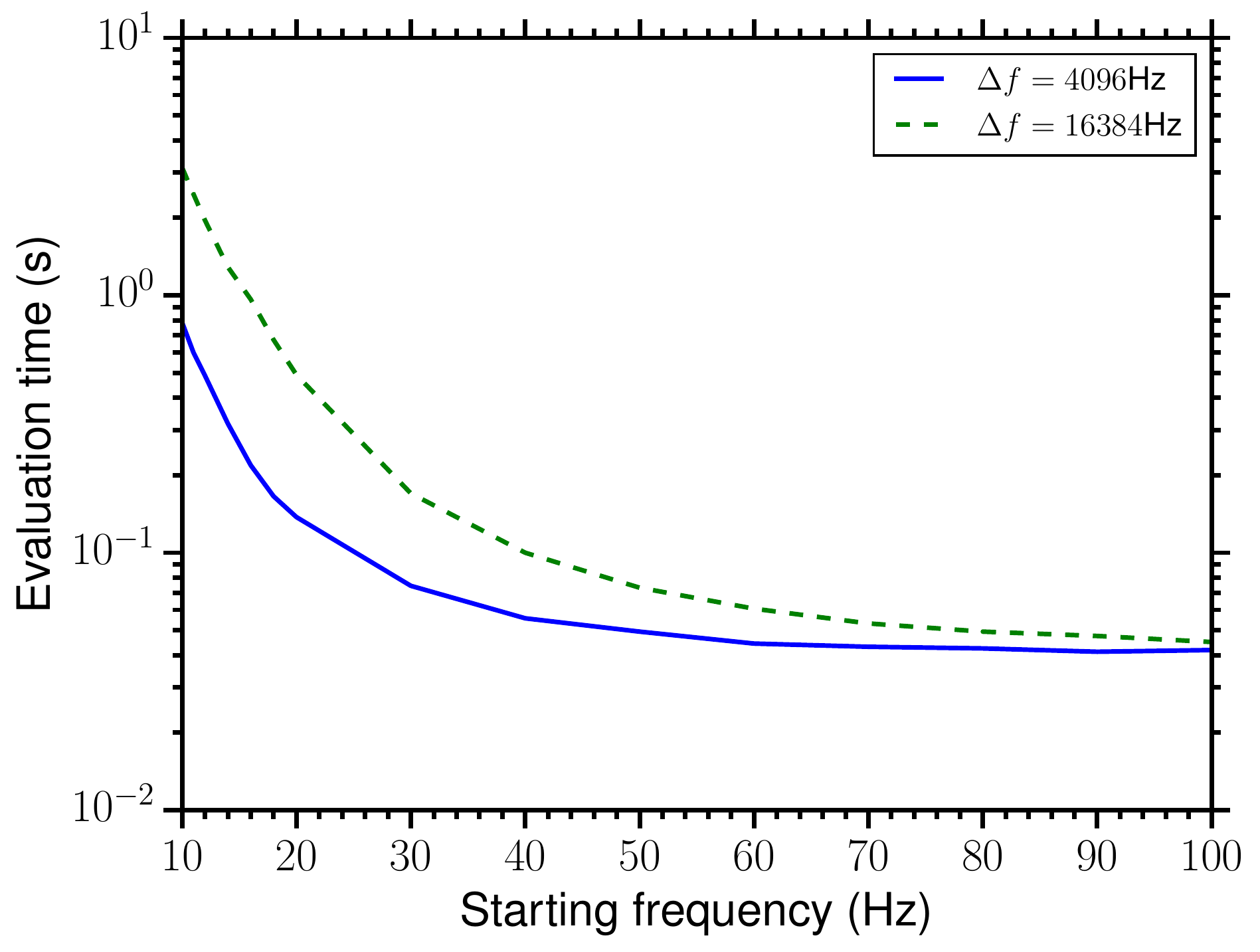}
\caption{Performance of the surrogate implemented in \texttt{LAL} for a
  binary with component masses $(1.4, 1.4)M_\odot$. We used
  sampling frequencies of 4096Hz and 16384Hz. The evaluation time at
  each starting frequency is averaged over 16 evaluations on a
  2.7~GHz Intel Xeon CPU using one core.}
\label{fig:timing}
\end{center}
\end{figure}

We have produced a prototype Python implementation of our surrogate and a C implementation in the 
LIGO Algorithm Library (LAL)~\cite{lal} under the name \texttt{\TEOBResum{}\_ROM} which is about 2 
times faster than the Python version. The performance of the LAL implementation is shown in Fig.~\ref{fig:timing}. 
For each waveform evaluation, there is a flat cost of $\sim 0.04$~s to calculate the amplitude and phase of 
the waveform at the $\sim 7\times10^4$ samples using Eq.~\eqref{eq:surrogate1}. The difference in evaluation time as the starting frequency 
$f_{\rm start}$ is varied results from the resampling of the amplitude and phase at evenly spaced times in Eq.~\eqref{eq:surrogate}. 
The number of these samples increases rapidly as the starting frequency is decreased. Most parameter 
estimation is done at 4096~Hz with a starting frequency of 30~Hz down to 10~Hz. The total waveform 
evaluation time will therefore be about 0.07~s up to 0.8~s. If necessary, the resampling of the amplitude and 
phase, which is the bottleneck for low starting frequencies, can be parallelized. The original implementation 
of the EOB code in Matlab can take $\sim 20$ minutes to evaluate, so this represents a speed-up factor 
of $10^3$--$10^4$ in some cases. However, it is not possible to make a direct comparison since the original
Matlab code calculates a non-uniformly sampled waveform with units rescaled by the distance and mass.

\subsection{Parameter estimation}

As an end-to-end test of the surrogate, we inject a waveform into simulated aLIGO data and estimate
its parameters using a Nested Sampling algorithm as implemented in 
LAL~\cite{Veitch:2008ur,Veitch:2008wd,Veitch:2009hd,Veitch:2014wba}. 
Although EOB waveforms have been used in Bayesian parameter estimation for BBH systems which
have significantly fewer cycles (e.g.~\cite{Abbott2016precession}), this is one of the first 
times EOB waveforms have been used in a Bayesian analysis for BNS systems. 
Another analysis using a parallelized algorithm~\cite{PankowBradyOchsner2015} is in 
progress~\cite{OshaughnessyRizzo2016}.

The synthetic data are taken 
to be stationary and Gaussian, with a PSD corresponding to the aLIGO final
design sensitivity \cite{Aasi:2013wya}; two detectors are assumed, located at the Hanford and
Livingston sites. For the simulated signal we choose an arbitrary sky position and orientation, 
a distance of 87 Mpc, and component masses $(M_A, M_B) = (1.4, 1.4)\,M_\odot$. The EOS is 
taken to be MS1b, so that for the given masses one has $\Lambda_2^A = \Lambda_2^B = 1286$ 
\cite{LackeyWade2015}. For the parameter choices made, the optimal signal-to-noise ratio
is 26.28. 

For the parameter estimation, the prior densities for sky position as well as orientation are chosen to 
be uniform on the sphere, and the distance prior is uniform in co-moving volume with an 
upper cut-off at 100 Mpc. Component masses are uniform in the interval $[1,2]\,M_\odot$, and 
we take $\Lambda_2^A$, $\Lambda_2^B$ to be uniform in the interval $[50,5000]$. In the 
Nested Sampling we use 512 live points and up to 5000 MCMC points \cite{Veitch:2009hd}, leading to 
$10^7 - 10^8$ likelihood evaluations; results from 4 different sampling chains are combined. Since the 
tidal effects we are interested in manifest themselves predominantly at high frequency, for this first 
exploration we use a lower cut-off frequency of 40 Hz; the sampling rate is 4096 Hz. 

Since $\Lambda_2^A$ and $\Lambda_2^B$ are highly correlated, after the 
Nested Sampling algorithm has finished we change parameters to the $\tilde{\Lambda}$, 
$\delta\tilde{\Lambda}$ introduced in \cite{WadeCreightonOchsner2014}, which depend on 
$(q, \Lambda_2^A, \Lambda_2^B)$; note that these have the convenient properties 
$\tilde{\Lambda}(q = 1, \Lambda_2^A = \Lambda_2^B = \Lambda) = \Lambda$ and 
$\delta\tilde{\Lambda}(q = 1, \Lambda_2^A = \Lambda_2^B = \Lambda) = 0$. 

\begin{figure}[t]
  \centering 
    \includegraphics[width=0.49\textwidth]{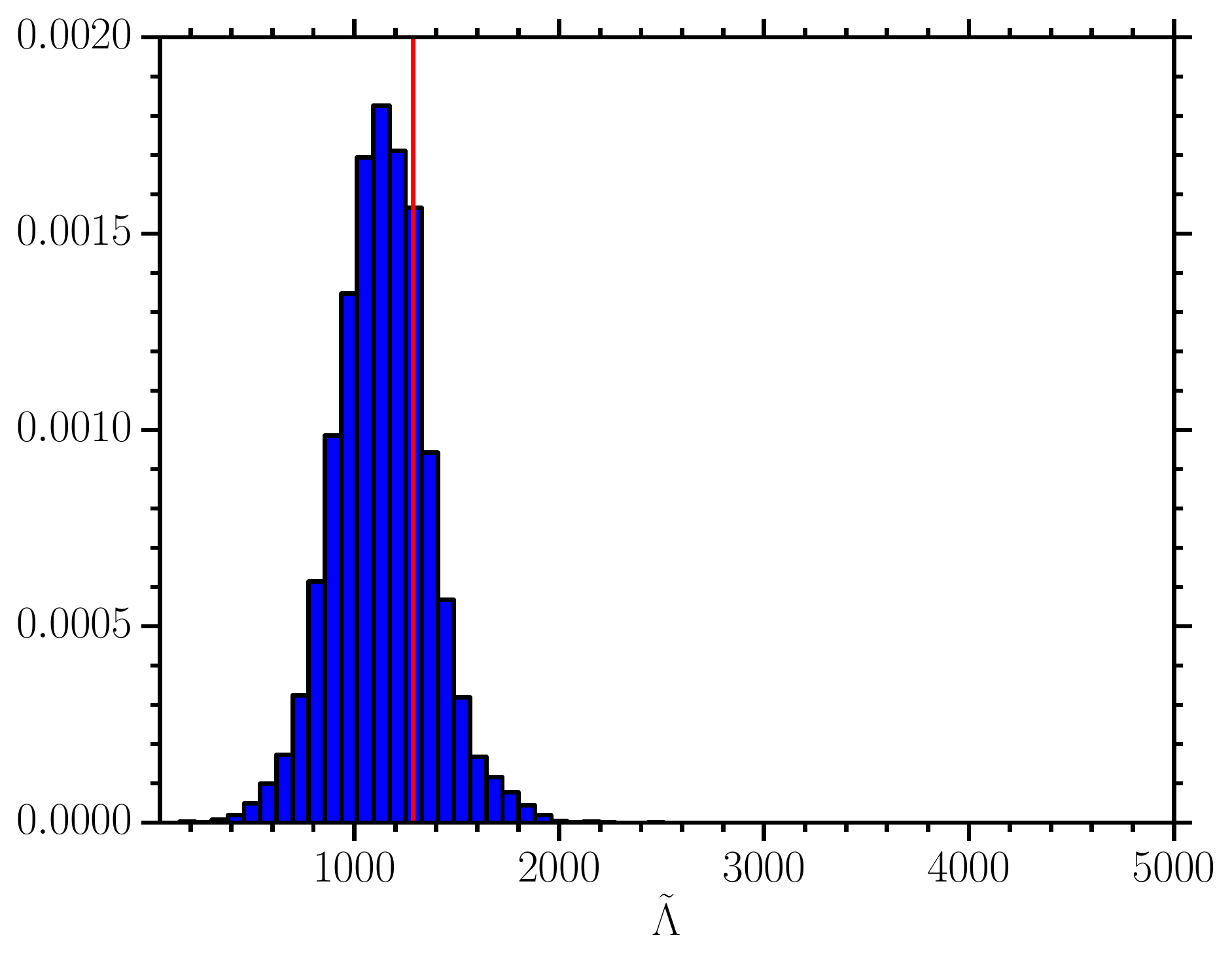}
    \includegraphics[width=0.49\textwidth]{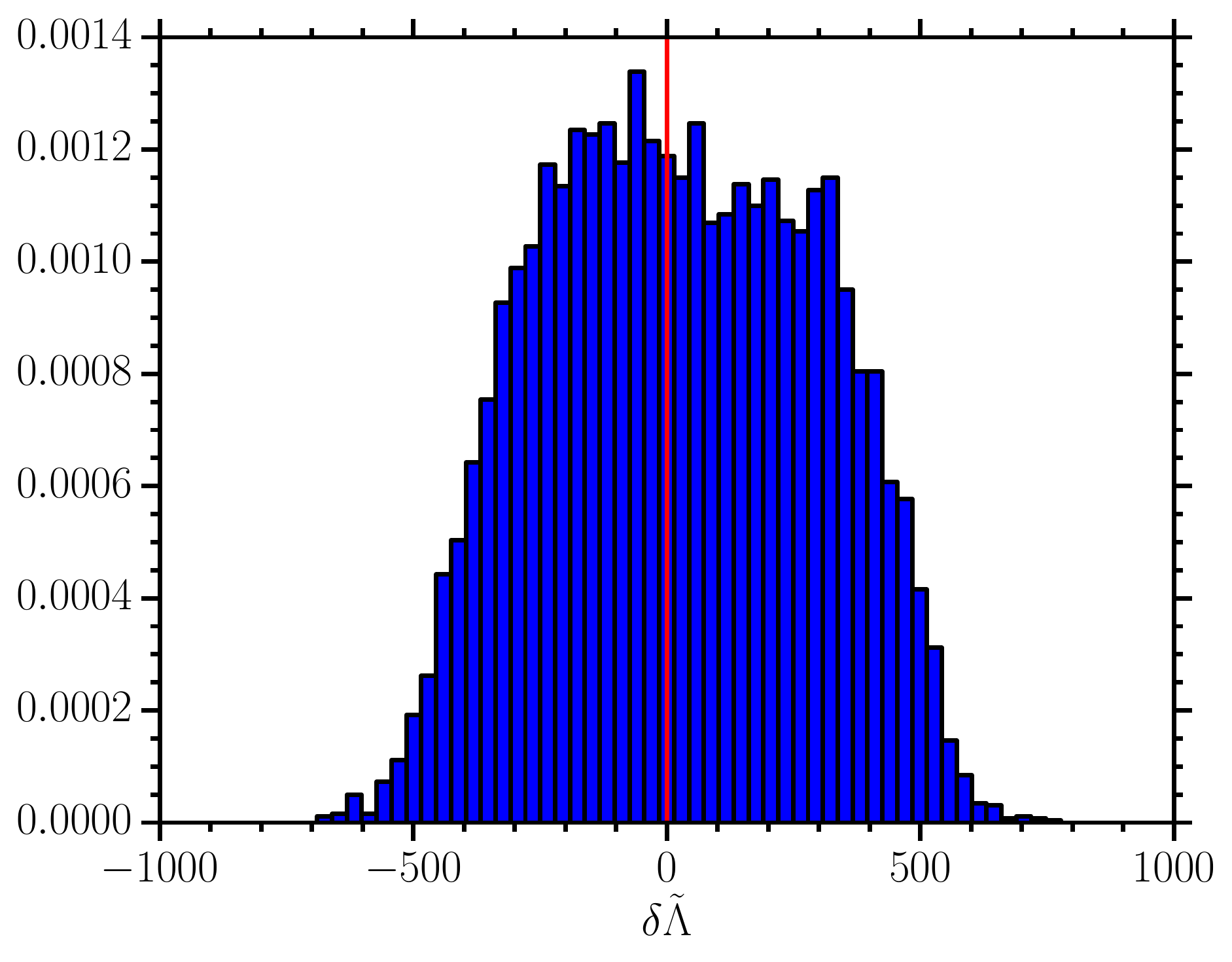}
    \caption{Posterior densities for the tidal parameters 
    $\tilde{\Lambda}$ (top) and $\delta\tilde{\Lambda}$ (bottom). The simulated signal had component 
    masses $(1.4, 1.4)\,M_\odot$ and EOS MS1b, so that $\tilde{\Lambda} = 1286$ and
    $\delta\tilde{\Lambda} = 0$, as indicated by the vertical red lines.}
       \label{fig:pe}
\end{figure}

Posterior density functions for $\tilde{\Lambda}$ and 
$\delta\tilde{\Lambda}$ are shown in Fig.~\ref{fig:pe}; both these parameters
are recovered quite well. The parameter estimation code ran for 15 days; 
without a surrogate it would have taken well over a year.

Finally, as shown in \cite{DelPozzoLiAgathos2013,AgathosMeidamDelPozzo2015,LackeyWade2015}, 
if the functional dependence of $\Lambda_2^A$, $\Lambda_2^B$ on component masses  
is expressed in terms of observables that can be expected to take approximately the same 
values for all sources, then posterior density functions for the latter can be trivially combined across
detections to arrive at a more accurate measurement. These observables could be coefficients
in a Taylor expansion \cite{DelPozzoLiAgathos2013,AgathosMeidamDelPozzo2015}, 
or the parameters in a representation of the EOS in terms of piecewise polytropes 
\cite{LackeyWade2015}. An implementation for the surrogate model presented in this paper
is left for future work.

\section{Discussion and future work}
\label{sec:theend}

We have constructed a surrogate for waveforms from nonspinning BNS systems that includes the $\ell=2$ tidal 
interaction and approximates the $\ell=3$ and 4 tidal interactions. The error of the surrogate is 
small compared to the size of the tidal effect, and we have demonstrated that this surrogate 
can be used in generic parameter estimation algorithms. 

The implementation of the model in LAL takes $\sim 0.07$~s to evaluate for $f_{\rm low} = 30$~Hz 
and $\sim 0.8$~s to evaluate for $f_{\rm low} = 10$~Hz, but there remains plenty of room for improvements. 
For starting frequencies below $\sim 30$~Hz, the waveform generation time is dominated by 
straightforward interpolation of the amplitude and phase at uniformly sampled times. This 
interpolation can be optimized or even parallelized on either CPUs or a GPU if necessary. 
Most other operations can be parallelized 
as well. In addition, the number of amplitude and phase bases as well as the the number of time 
samples used to store the amplitude and phase bases (currently $\sim 7 \times 10^4$) can be 
optimized, and the number of Chebyshev coefficients for the interpolation can be strategically 
reduced as well. We expect speed-up factors of a few should be possible without significantly 
effecting the accuracy of the surrogate. Finally, one could use our time domain surrogate to 
build a linear frequency domain surrogate that can be used directly in a reduced-order 
quadratures implementation of likelihood computations for speeding up parameter estimation studies.

Because EOB models that include both spin and tidal interactions are still in 
progress~\cite{Bernuzzi2016, HindererSteinhoffTaracchini2016}, our surrogate leaves out spin parameters. Once 
these models are available, adding the two spin magnitudes $|S_1|$ and $|S_2|$ for aligned 
spin systems for a total of five parameters will likely be straightforward using 
a standard grid-based interpolation scheme.
However, incorporating an additional four parameters to 
account for the spin orientations will likely be significantly more difficult. So far this problem 
has not been fully solved for BBH systems without tidal interactions.

Finally, we have also left out the post-merger stage for BNS systems. Unlike BBH systems, the 
post-merger stage can only be modeled by expensive numerical relativity simulations, and it is 
unlikely that more than 100--1000 simulations could be performed over the course of a few years. 
However, work by Clark {\it et al}. has shown that it is possible to reconstruct post-merger waveforms 
with a small number of orthonormal bases~\cite{ClarkBausweinStergioulas2016}. In future work, 
we would like to examine the possibility of constructing a surrogate model for the complete 
inspiral--post-merger waveform for BNS systems.

\begin{acknowledgments}

BL thanks Larne Pekowsky and Duncan Brown for significant computing help 
and Rory Smith and Michael P{\"u}rrer for helpful discussions
at the beginning of this work. 
SB thanks Paolo Pani for helpful discussions about
$\Lambda^{\rm fit}_{3,4}(\Lambda_2)$ fits. 
BL was supported by NSF grant AST-1333142.
SB was supported by a Rita Levi Montalcini fellowship of the Italian Ministry of Education,
University and Research (MIUR).
CRG was supported in part by NSF grant PHY-1404569 to Caltech and by the Sherman Fairchild Foundation. 
JM and CVDB were supported by the research programme of the Foundation for Fundamental 
Research on Matter (FOM), which is partially supported by the Netherlands Organisation 
for Scientific Research (NWO). 
Computations were performed on the Syracuse University Campus Grid which is supported 
by NSF awards ACI-1341006, ACI-1541396, and by Syracuse University ITS.

\end{acknowledgments}

\appendix

\section{Systematic uncertainties related to $\ell=3,4$ tidal
  polarizability coefficients}
\label{app:Yagifits}

\begin{figure}[t]
  \centering 
    \includegraphics[width=0.49\textwidth]{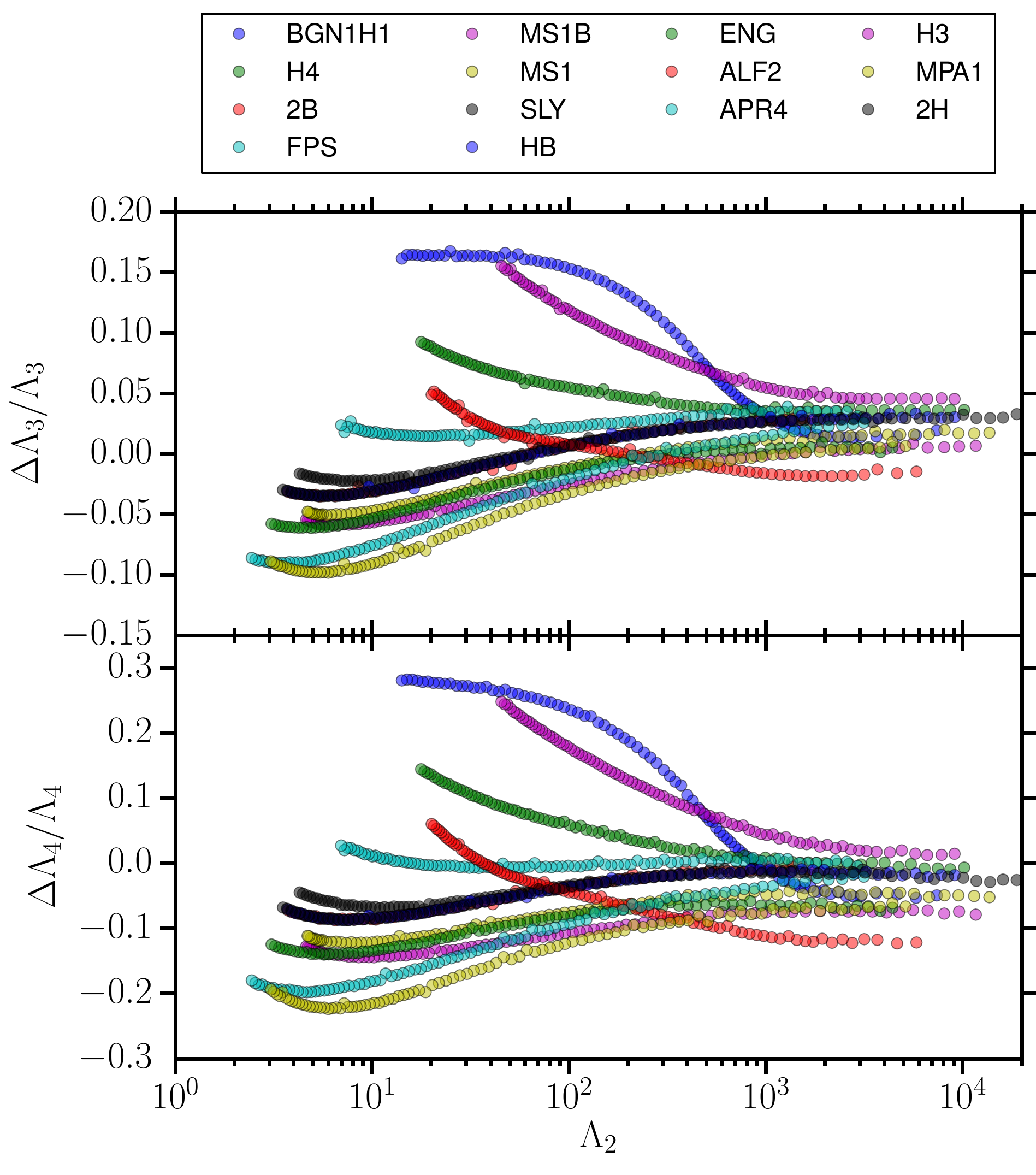}
    \caption{Relative errors $\Delta\Lambda_{3}/\Lambda_{3}$ and $\Delta\Lambda_{4}/\Lambda_{4}$, with
      $\Delta\Lambda_\ell=\Lambda_\ell-\Lambda_\ell^\text{fit}$ (shown only 300 points).}
 \label{fig:error_Yagifits}
\end{figure}

In our 3D surrogate the quantities $\Lambda_{3,4}$ are computed from $\Lambda_2$ using \cite{Yagi:2013sva}
\begin{equation}
\label{eq:lamda34fit}
\ln \Lambda^\text{fit}_{\l} = \sum_{i=0}^4 a^{(\l)}_i (\ln \Lambda_2)^i \ \ (\l=3,4)
\end{equation}
where $(a^{(3)}_i) = (-1.15, 1.18, 2.51\times10^{-2},
-1.31\times10^{-3}, 2.52\times10^{-5})$ and $(a^{(4)}_i)
= (-2.45, 1.43, 3.95\times10^{-2}, -1.81\times10^{-3},
2.80\times10^{-5})$. 
The accuracy of these fits and the systematic uncertainties that
they introduce in the surrogate are tested using a sample of 14 EOS and,
for each EOS, about 3000 star configurations spanning the mass range
$M\in[0.9,M_\text{max}]M_\odot$. 

Figure~\ref{fig:error_Yagifits} and Table~\ref{tab:fit_uncert} quantify the relative errors of the
fits, where by definition $\Delta\Lambda_\ell=\Lambda_\ell-\Lambda_\ell^\text{fit}$. 
The largest fit errors occur for small $\Lambda_2\sim1-10$, which occur when the NS is near its maximum mass.
The largest positive error is for EOS BGN1H1 near its maximum mass of $1.64M_\odot$ and the 
largest negative error is for MPA1 near its maximum mass of $2.43M_\odot$. These fit errors have the range 
$\Delta\Lambda_3/\Lambda_3 \in [-0.10, 0.17]$ and $\Delta\Lambda_4/\Lambda_4 \in [-0.22, 0.29]$.
These high mass configurations, however, are not expected to be found in a BNS system. Comparing with 
Ref.~\cite{Yagi:2013sva}, we obtain a larger range of errors because we use a larger sample of EOS.

Because the tidal effect is largest for equal mass systems and the tidal parameter is largest for smaller masses, 
the error in the waveform phase due to the fit is largest for equal mass $q=1$ systems with smaller masses. 
In Fig.~\ref{fig:eosphase} and Table~\ref{tab:fit_uncert}, we show the phase error from
NSs with more likely masses of $1.4M_\odot$ using the soft EOS SLY and the stiff EOS MS1b. 
The phase error grows with time, reaching it's maximum near merger, and in general, we find typical phase 
errors of $|\Delta\phi| \lesssim 0.01$~radians.

\begin{table*}[t]
  \centering    
  \caption{Deviations in waveform phase due to errors in the fits $\Lambda^{\rm fit}_3(\Lambda_2)$ and $\Lambda^{\rm fit}_4(\Lambda_2)$. 
  Waveforms are calculated with a mass ratio of $q=1$. Phase differences are computed as 
  $|\Delta\phi(t)|=|\phi(t)-\phi^\text{fit}(t)|$ and the maximum difference up to the amplitude peak of the shorter waveform is listed.} 
    \begin{tabular}{cccccccc}        
      \hline
      EOS & M ($M_\odot$) & $\Lambda_2$ & $\Lambda_3$ & $\Delta\Lambda_3/\Lambda_3$ & $\Lambda_4$ & $\Delta\Lambda_4/\Lambda_4$ & $|\Delta\phi|$ \\
    \hline
    BGN1H1 & 1.64 & 14 & 9.7 & 0.17 & 6.5 & 0.29 & 0.002 \\
    MPA1 & 2.47 & 3.05 & 1.1 & -0.089 & 0.37 & -0.19 & 0.0005 \\
    SLY & 1.4 & 307 & 511 & 0.024 & 819 & -0.017 & 0.003 \\
    MS1b & 1.4 & 1260 & 3440 & 0.0001 & 9080 & -0.075 & 0.007 \\
     \hline
  \end{tabular}
 \label{tab:fit_uncert}
\end{table*}

\bibliography{paper,rb,inspire-hep}  

\end{document}